\documentclass[aps,prd,eqsecnum,showpacs,superscriptaddress,
twocolumn,nofootinbib,floatfix,preprintnumbers,altaffilletter]{revtex4}
\usepackage{natbib}
\usepackage[caption=false]{subfig}
\usepackage{pdfpages}
\usepackage{graphicx}
\usepackage{hyperref}
\usepackage{amssymb}
\usepackage{amsmath}
\usepackage{longtable}
\usepackage{rotating}
\usepackage{color}
\usepackage{fancyhdr}

\def\lesssim{\mathrel{\hbox{\rlap{\hbox{\lower4pt\hbox{$\sim$}}}\hbox{$<$}}}}
\def\gtrsim{\mathrel{\hbox{\rlap{\hbox{\lower4pt\hbox{$\sim$}}}\hbox{$>$}}}}

\newcommand{\subf}[2]{%
  {\small\begin{tabular}[t]{@{}c@{}}
  #1\\#2
  \end{tabular}}%
}

\newcommand{\private}[1]{}

\newcommand{\be}{\begin{equation}}
\newcommand{\ee}{\end{equation}}
\newcommand{\bea}{\begin{eqnarray}}
\newcommand{\eea}{\end{eqnarray}}
\newcommand{\bdm}{\begin{displaymath}}
\newcommand{\edm}{\end{displaymath}}

\begin{document}
\title{Experimental results from the ST7 mission on LISA Pathfinder}

\def\addressa{European Space Astronomy Centre, European Space Agency, Villanueva de la Ca\~{n}ada, 28692 Madrid, Spain}
\def\addressb{Albert-Einstein-Institut, Max-Planck-Institut f\"ur Gravitationsphysik und Leibniz Universit\"at Hannover, Callinstra{\ss}e 38, 30167 Hannover, Germany}
\def\addressc{APC, Univ Paris Diderot, CNRS/IN2P3, CEA/lrfu, Obs de Paris, Sorbonne Paris Cit\'e, France}
\def\addressd{High Energy Physics Group, Physics Department, Imperial College London, Blackett Laboratory, Prince Consort Road, London, SW7 2BW, UK }
\def\addresse{Dipartimento di Fisica, Universit\`a di Roma ``Tor Vergata'',  and INFN, sezione Roma Tor Vergata, I-00133 Roma, Italy}
\def\addressf{Department of Industrial Engineering, University of Trento, via Sommarive 9, 38123 Trento, and Trento Institute for Fundamental Physics and Application / INFN}
\def\addressh{European Space Technology Centre, European Space Agency, Keplerlaan 1, 2200 AG Noordwijk, The Netherlands}
\def\addressi{Dipartimento di Fisica, Universit\`a di Trento and Trento Institute for Fundamental Physics and Application / INFN, 38123 Povo, Trento, Italy}
\def\addressk{Istituto di Fotonica e Nanotecnologie, CNR-Fondazione Bruno Kessler, I-38123 Povo, Trento, Italy}
\def\addressj{The School of Physics and Astronomy, University of Birmingham, Birmingham, UK}
\def\addressl{Institut f\"ur Geophysik, ETH Z\"urich, Sonneggstrasse 5, CH-8092, Z\"urich, Switzerland}
\def\addressm{The UK Astronomy Technology Centre, Royal Observatory, Edinburgh, Blackford Hill, Edinburgh, EH9 3HJ, UK}
\def\addressn{Institut de Ci\`encies de l'Espai (CSIC-IEEC), Campus UAB, Carrer de Can Magrans s/n, 08193 Cerdanyola del Vall\`es, Spain}
\def\addresso{DISPEA, Universit\`a di Urbino ``Carlo Bo'', Via S. Chiara, 27 61029 Urbino/INFN, Italy}
\def\addressp{European Space Operations Centre, European Space Agency, 64293 Darmstadt, Germany }
\def\addressq{Physik Institut, Universit\"at Z\"urich, Winterthurerstrasse 190, CH-8057 Z\"urich, Switzerland}
\def\addressr{SUPA, Institute for Gravitational Research, School of Physics and Astronomy, University of Glasgow, Glasgow, G12 8QQ, UK}
\def\addresss{Department d'Enginyeria Electr\`onica, Universitat Polit\`ecnica de Catalunya,  08034 Barcelona, Spain}
\def\addresst{Institut d'Estudis Espacials de Catalunya (IEEC), C/ Gran Capit\`a 2-4, 08034 Barcelona, Spain}
\def\addressgsfc{Gravitational Astrophysics Lab, NASA Goddard Space Flight Center, 8800 Greenbelt Road, Greenbelt, MD 20771 USA}
\def\addressbb{Department of Mechanical and Aerospace Engineering, MAE-A, P.O. Box 116250, University of Florida, Gainesville, Florida 32611, USA}
\def\addresscc{Istituto di Fotonica e Nanotecnologie, CNR-Fondazione Bruno Kessler, I-38123 Povo, Trento, Italy}
\def\addressjpl{NASA Jet Propulsion Laboratory, California Institute of Technology, Pasadena, CA 91109 USA}
\def\addressham{The Hammers Co., Greenbelt, MD 20771 USA}
\def\addressbus{Busek Co., Natick, MA 01760 USA}



 \author{G~Anderson}\affiliation{\addressjpl}
 \author{J~Anderson}\affiliation{\addressjpl}
 \author{M~Anderson}\affiliation{\addressjpl}
 \author{G~Aveni}\affiliation{\addressjpl}
 \author{D~Bame}\affiliation{\addressjpl}
 \author{P~Barela}\affiliation{\addressjpl}
  \author{K~Blackman}\affiliation{\addressham}
   \author{A~Carmain}\affiliation{\addressjpl}
   \author{L~Chen}\affiliation{\addressjpl}
 \author{M~Cherng}\affiliation{\addressjpl}
 \author{S~Clark}\affiliation{\addressjpl}
 \author{M~Connally}\affiliation{\addressjpl}
 \author{W~Connolly}\affiliation{\addressbus}
 \author{D~Conroy}\affiliation{\addressjpl}
 \author{M~Cooper}\affiliation{\addressjpl}
 \author{C~Cutler}\affiliation{\addressjpl}
 \author{J~D'Agostino}\affiliation{\addressham}
 \author{N~Demmons}\affiliation{\addressbus}
  \author{E~Dorantes}\affiliation{\addressjpl}
 \author{C~Dunn}\affiliation{\addressjpl}
 \author{M~Duran}\affiliation{\addressjpl}
\author{E~Ehrbar}\affiliation{\addressbus}
 \author{J~Evans}\affiliation{\addressjpl}
 \author{J~Fernandez}\affiliation{\addressjpl}
  \author{G~Franklin}\affiliation{\addressjpl}
   \author{M~Girard}\affiliation{\addressjpl}
    \author{J~Gorelik}\affiliation{\addressjpl}
     \author{V~Hruby}\affiliation{\addressbus}
 \author{O~Hsu}\affiliation{\addressgsfc}
  \author{D~Jackson}\affiliation{\addressjpl}
 \author{S~Javidnia}\affiliation{\addressjpl}
  \author{D~Kern}\affiliation{\addressjpl}
 \author{M~Knopp}\affiliation{\addressjpl}
 \author{R~Kolasinski}\affiliation{\addressjpl}
  \author{C~Kuo}\affiliation{\addressjpl}
   \author{T~Le}\affiliation{\addressjpl}
    \author{I~Li}\affiliation{\addressjpl}
 \author{O~Liepack}\affiliation{\addressjpl}
 \author{A~Littlefield}\affiliation{\addressjpl}
 \author{P~Maghami}\affiliation{\addressgsfc}
 \author{S~Malik}\affiliation{\addressjpl}
  \author{L~Markley}\affiliation{\addressgsfc}
 \author{R~Martin}\affiliation{\addressbus}
  \author{C~Marrese-Reading}\affiliation{\addressjpl}
   \author{J~Mehta}\affiliation{\addressjpl}
 \author{J~Mennela}\affiliation{\addressjpl}
 \author{D~Miller}\affiliation{\addressjpl}
  \author{D~Nguyen}\affiliation{\addressjpl}
   \author{J~O'Donnell}\affiliation{\addressgsfc}
    \author{R~Parikh}\affiliation{\addressjpl}
     \author{G~Plett}\affiliation{\addressjpl}
      \author{T~Ramsey}\affiliation{\addressjpl}
 \author{T~Randolph}\affiliation{\addressjpl}
  \author{S~Rhodes}\affiliation{\addressbus}
 \author{A~Romero-Wolf}\affiliation{\addressjpl}
 \author{T~Roy}\affiliation{\addressbus}
 \author{A~Ruiz}\affiliation{\addressjpl}
  \author{H~Shaw}\affiliation{\addressjpl}
   \author{J~Slutsky}\affiliation{\addressgsfc}
   \author{D~Spence}\affiliation{\addressbus}
  \author{J~Stocky}\affiliation{\addressjpl}
 \author{J~Tallon}\affiliation{\addressjpl}
  \author{I~Thorpe}\affiliation{\addressgsfc}
 \author{W~Tolman}\affiliation{\addressbus}
  \author{H~Umfress}\affiliation{\addressjpl}
   \author{R~Valencia}\affiliation{\addressjpl}
 \author{C~Valerio}\affiliation{\addressjpl}
 \author{W~Warner}\affiliation{\addressjpl}
 \author{J~Wellman}\affiliation{\addressjpl}
 \author{P~Willis}\affiliation{\addressjpl}
 \author{J~Ziemer}\affiliation{\addressjpl}
 \author{J~Zwahlen}\affiliation{\addressbus}
\collaboration{The ST7 Team}
\noaffiliation


\author{M~Armano}\affiliation{\addressh}
\author{H~Audley}\affiliation{\addressb}
\author{J~Baird}\affiliation{\addressc}
\author{P~Binetruy}\thanks{Deceased 30 March 2017}\affiliation{\addressc}
\author{M~Born}\affiliation{\addressb}
\author{D~Bortoluzzi}\affiliation{\addressf}
\author{E~Castelli}\affiliation{\addressi}
\author{A~Cavalleri}\affiliation{\addresscc}
\author{A~Cesarini}\affiliation{\addresso}
\author{A\,M~Cruise}\affiliation{\addressj}
\author{K~Danzmann}\affiliation{\addressb}
\author{M~de Deus Silva}\affiliation{\addressa}
\author{I~Diepholz}\affiliation{\addressb}
\author{G~Dixon}\affiliation{\addressj}
\author{R~Dolesi}\affiliation{\addressi}
\author{L~Ferraioli}\affiliation{\addressl}
\author{V~Ferroni}\affiliation{\addressi}
\author{E\,D~Fitzsimons}\affiliation{\addressm}
\author{M~Freschi}\affiliation{\addressa}
\author{L~Gesa}\affiliation{\addressn}
\author{F~Gibert}\affiliation{\addressi}
\author{D~Giardini}\affiliation{\addressl}
\author{R~Giusteri}\affiliation{\addressi}
\author{C~Grimani}\affiliation{\addresso}
\author{J~Grzymisch}\affiliation{\addressh}
\author{I~Harrison}\affiliation{\addressp}
\author{G~Heinzel}\affiliation{\addressb}
\author{M~Hewitson}\affiliation{\addressb}
\author{D~Hollington}\affiliation{\addressd}
\author{D~Hoyland}\affiliation{\addressj}
\author{M~Hueller}\affiliation{\addressi}
\author{H~Inchausp\'e}\affiliation{\addressc}\affiliation{\addressbb}
\author{O~Jennrich}\affiliation{\addressh}
\author{P~Jetzer}\affiliation{\addressq}
\author{N~Karnesis}\affiliation{\addressc}
\author{B~Kaune}\affiliation{\addressb}
\author{N~Korsakova}\affiliation{\addressr}
\author{C\,J~Killow}\affiliation{\addressr}
\author{J\,A~Lobo}\thanks{Deceased 30 September 2012}\affiliation{\addressn}
\author{I~Lloro}\affiliation{\addressn}
\author{L~Liu}\affiliation{\addressi}
\author{J\,P~L\'opez-Zaragoza}\affiliation{\addressn}
\author{R~Maarschalkerweerd}\affiliation{\addressp}
\author{D~Mance}\affiliation{\addressl}
\author{N~Meshksar}\affiliation{\addressl}
\author{V~Mart\'{i}n}\affiliation{\addressn}
\author{L~Martin-Polo}\affiliation{\addressa}
\author{J~Martino}\affiliation{\addressc}
\author{F~Martin-Porqueras}\affiliation{\addressa}
\author{I~Mateos}\affiliation{\addressn}
\author{P\,W~McNamara}\affiliation{\addressh}
\author{J~Mendes}\affiliation{\addressp}
\author{L~Mendes}\affiliation{\addressa}
\author{M~Nofrarias}\affiliation{\addressn}
\author{S~Paczkowski}\affiliation{\addressb}
\author{M~Perreur-Lloyd}\affiliation{\addressr}
\author{A~Petiteau}\affiliation{\addressc}
\author{P~Pivato}\affiliation{\addressi}
\author{E~Plagnol}\affiliation{\addressc}
\author{J~Ramos-Castro}\affiliation{\addresss}
\author{J~Reiche}\affiliation{\addressb}
\author{D\,I~Robertson}\affiliation{\addressr}
\author{F~Rivas}\affiliation{\addressn}
\author{G~Russano}\affiliation{\addressi}
\author{C\,F~Sopuerta}\affiliation{\addressn}
\author{T~Sumner}\affiliation{\addressd}
\author{D~Texier}\affiliation{\addressa}
\author{D~Vetrugno}\affiliation{\addressi}
\author{S~Vitale}\affiliation{\addressi}
\author{G~Wanner}\affiliation{\addressb}
\author{H~Ward}\affiliation{\addressr}
\author{P\,J~Wass}\affiliation{\addressd}\affiliation{\addressbb}
\author{W\,J~Weber}\affiliation{\addressi}
\author{L~Wissel}\affiliation{\addressb}
\author{A~Wittchen}\affiliation{\addressb}
\author{P~Zweifel}\affiliation{\addressl}
\collaboration{The LISA Pathfinder Collaboration}
\noaffiliation

\date{\today}

\begin{abstract}
The Space Technology 7 Disturbance Reduction System (ST7-DRS) is a NASA technology demonstration payload that operated from January 2016 through July of 2017 on the European Space Agency's LISA Pathfinder spacecraft .  The joint goal of the NASA and ESA missions was to validate key technologies for a future space-based gravitational wave observatory targeting the source-rich milliHertz band. The two primary components of ST7-DRS are a micropropulsion system based on colloidal micro-Newton thrusters (CMNTs) and a control system that simultaneously controls the attitude and position of the spacecraft and the two free-flying test masses (TMs).  This paper presents our main experimental results and summarizes the overall the performance of the CMNTs and control laws. We find that the CMNT performance to be consistent with pre-flight predictions, with a measured system thrust noise on the order of $100\,\textrm{nN}/\sqrt{\textrm{Hz}}$ in the $1\,\textrm{mHz}\leq f \leq 30\,\textrm{mHz}$ band. The control system maintained the TM-spacecraft separation with an RMS error of less than 2$\,$nm and a noise spectral density of less than $3\,\textrm{nm}/\sqrt{\textrm{Hz}}$ in the same band.  Thruster calibration measurements yield thrust values consistent with the performance model and ground-based thrust-stand measurements, to within a few percent. We also report a differential acceleration noise between the two test masses with a spectral density of roughly  $3\,\textrm{fm}/\textrm{s}^2/\sqrt{\textrm{Hz}}$ in the $1\,\textrm{mHz}\leq f \leq 30\,\textrm{mHz}$ band, slightly less than twice as large as the best performance reported with the baseline LISA Pathfinder configuration and below the current requirements for the Laser Interferometer Space Antenna (LISA) mission. 
\end{abstract}

\pacs{07.05.Dz, 07.87.+v}
\maketitle 
\section{Introduction}
\label{sec:INTRO}
\subsection{LISA Pathfinder, the LISA Technology Package, and ST7-DRS}

The Space Technology 7 Disturbance Reduction System (ST7-DRS) is a NASA technology demonstration payload hosted on the European Space Agency (ESA) LISA Pathfinder (LPF) spacecraft, which launched from Kourou, French Guiana on December 3, 2015 and operated until July 17th, 2017, when it was decommissioned by ESA.  The primary purpose of LPF was to validate key elements of the measurement concept for the Laser Interferometer Space Antenna (LISA), a planned space-based mission to observe gravitational waves in the millihertz band.  Specifically, LPF demonstrated that the technique of drag-free control could be employed to place a test mass in near-perfect free-fall~\cite{LPF_PRL_2016,LPF_PRL_2018}. LISA will use three drag-free satellites, configured as an equilateral triangle with $\sim 2.5$ million km arms, to detect spacetime strains caused by passing gravitational waves~\cite{LISA_PROPOSAL_2017}.

The basic components of a drag-free system are the reference test mass, which resides inside the spacecraft but makes no physical contact with it; a metrology system that measures the position and attitude of the test mass relative to the spacecraft as an inertial sensor; a control system that determines what forces and torques to apply to the spacecraft, and possibly the test mass; and an actuation system that can apply forces and torques to the spacecraft and possibly the test mass.  In the case of LPF, European National Space Agencies provided the LISA Technology Package (LTP), which includes two test masses as part of the inertial sensor.  Each test mass has its own independent six-degree-of-freedom electrostatic metrology and control system.  LTP also includes an optical interferometer that measures the position and attitude of the test masses with respect to the spacecraft and each other much more precisely than the electrostatic system, but only along the axis that joins the two test masses as well as the tip and tilt angles orthogonal to that axis. Finally, the LTP includes systems to monitor and control the thermal, magnetic, and charge environment of the instrument.  The ESA-provided spacecraft included its own set of drag-free control laws and its own cold-gas micropropulsion system.  ESA's drag-free system was used for the majority of LPF's operations and achieved a striking level of performance, significantly exceeding the requirements set for LPF (which were deliberately relaxed from the LISA requirements) and meeting or exceeding the requirements for LISA itself~\cite{LPF_PRL_2016,LPF_PRL_2018}. 
 
ST7-DRS includes two main elements: an alternate set of drag-free control laws implemented on a separate computer, and an alternate micropropulsion system based on a novel colloidal microthruster technology~\cite{Ziemer2006, Ziemer2008}. ST7-DRS provided the first demonstration of colloidal micropropulsion performance in space.   During phases of the LPF mission where ST7-DRS operated,  NASA's colloidal thrusters were used in place of ESA's cold-gas thrusters to move and orient the spacecraft, with the DRS control laws replacing the ESA control laws. For brief periods, NASA's colloid thrusters were also used as the actuators for ESA's drag-free system, replacing the cold gas thrusters, to show performance and the robust nature of the drag-free control laws and colloid microthruster technology. 
 
During ST7 operations, the LTP payload played the same role as during the ESA-led parts of the mission - providing information on the positions and attitudes of the test masses and applying forces and torques to the test masses, as requested by the DRS controllers. In this paper, we present an overview of the ST7-DRS operations, the measured performance of the ST7-DRS systems, and the implications for LISA.
  
\subsection{History of ST7-DRS development and operations}
 
Initiated in 2002 as part of NASA's New Millennium program,  ST7-DRS includes four subsystems: (1) The Integrated Avionics Unit (IAU), a computer based on the RAD750 processor; (2) Colloid Micro Newton Thrusters (CMNT), two clusters of four thrusters each; (3) Dynamic Control Software (DCS), a software subsystem which implements drag-free control algorithms and (4) Flight Software (FSW), a command and data handling software subsystem which processes commands and telemetry and hosts the DCS. The IAU was manufactured by Broadreach Engineering (Phoenix, AZ) and was delivered to NASA's Jet Propulsion Laboratory (JPL) for integrated testing in May, 2006. The CMNTs were manufactured by Busek (Natick, MA), put through acceptance and thermal testing in late 2007, and delivered to JPL with fully loaded propellant tanks in 2008. The DCS software was written at NASA's Goddard Spaceflight Center (GSFC) and the FSW was written at JPL, with initial versions both completed in March, 2006. The DRS completed a Pre-Ship Acceptance Review with ESA in June 2008 and was placed in storage until its delivery to Astrium UK, in Stevenage, England, for Assembly, Integration, Verification and Test (AIVT) in July 2009.

Due to the the unexpectedly long duration between DRS delivery and LPF's launch, ST7 conducted shelf-life extension testing on the thruster propellant, materials, and microvalves in both 2010 and 2013, which qualified the system for launch in 2015 and serendipitously demonstrated a long storage lifetime (8 years on the ground and nearly 10 years total with on-orbit operations) that will be useful for LISA.  During the storage period, the thrusters were left on the spacecraft, fully loaded with propellant, with removable protective covers on each thruster head to prevent debris from entering the electrodes.  During this time, the spacecraft was kept in the integration and test facilities at Airbus Stevenage, UK with dynamic and thermal environmental testing occurring at IABG in Ottobrunn, Germany.  The thrusters were part of all spacecraft-level testing with at least annual inspections removing the protective covers, none of which showed any signs of propellant leakage or damage to the thruster electrodes.  The thrusters required no special handling or environmental control beyond the normal safeguards and environments used during typical spacecraft AIVT activities, and the protective covers were removed just before spacecraft encapsulation into the launch fairing.
 
After launch, a composite of the LPF Science Module (SCM) and Propulsion Module (PRM) executed a series of orbit raising maneuvers culminating in a cruise to Earth-Sun L1. DRS, which was powered-off during launch, was turned on for initial commissioning January 2 - 10,  2016.  Because the PRM was still fastened to SCM, the DRS did not control the spacecraft  attitude in this commissioning, but the effects of the DRS thrusters were observed in the host spacecraft attitude control using on-board gyroscopes. LPF arrived on station at L1 on January 22, 2016 and the PRM was was discarded leaving the SCM. At this time, the ESA LISA Technology Package (LTP) was commissioned and began executing its primary mission on March 1st, 2016.  A second commissioning of DRS was conducted June 27 - July 8, 2016,  which included successful demonstrations of drag-free control. DRS operations were conducted over the next five months including 13 different experiments with occasional breaks for planned LTP station-keeping maneuvers or LTP experiments, as well as response to a number of anomalies on both DRS and LPF hardware.  The DRS anomalies are discussed in Section~\ref{sec:ANOMALIES}. DRS completed its baseline mission on December 6, 2016. An extended mission to further characterize the thrusters and control system was requested and approved, and operated from Mar 17, 2017 to April 30, 2017.  The DRS was decommissioned as part of the LPF decommissioning process on July 13, 2017.  Key dates for DRS operations are summarized in Table \ref{tab:dates}.    The complete ST7-DRS data set, along with tools for accessing it, is archived at \url{https://heasarc.gsfc.nasa.gov/docs/lpf/}.

\begin{table}
\begin{tabular}{|p{0.9in}|p{0.6in}|p{0.9in}|p{0.6in}|}
\hline
Event & Date & Event & Date \\ \hline
LPF Launch & 03 Dec '15 & Thruster-4 Anomaly & 27 Oct '16 \\ \hline
Transfer Phase Commissioning (9 days) & 02 Jan '16 & Start:Hybrid Propulsion & 29 Nov '16 \\ \hline
Arrival at L1 & 22 Jan '16 & End:Primary Mission & 06 Dec '16 \\ \hline
Experiment Phase Commissioning (10 days) & 27 Jun '16 & Start:Extended Mission & 20 Mar '17 \\ \hline
Cluster-2 DCIU Anomaly & 09 Jul '16 & End:Extended Mission  & 30 Apr '17 \\ \hline
Start:Primary Mission & 15 Aug '16 & Decommissioning Activities & 13 Jul '17 \\ \hline
\end{tabular}
\caption{ Key dates for DRS operations}\label{tab:dates}
\end{table}
\normalsize

\subsection{DRS Components and Interfaces}

\begin{figure}
\begin{center}
\includegraphics[width=\columnwidth]{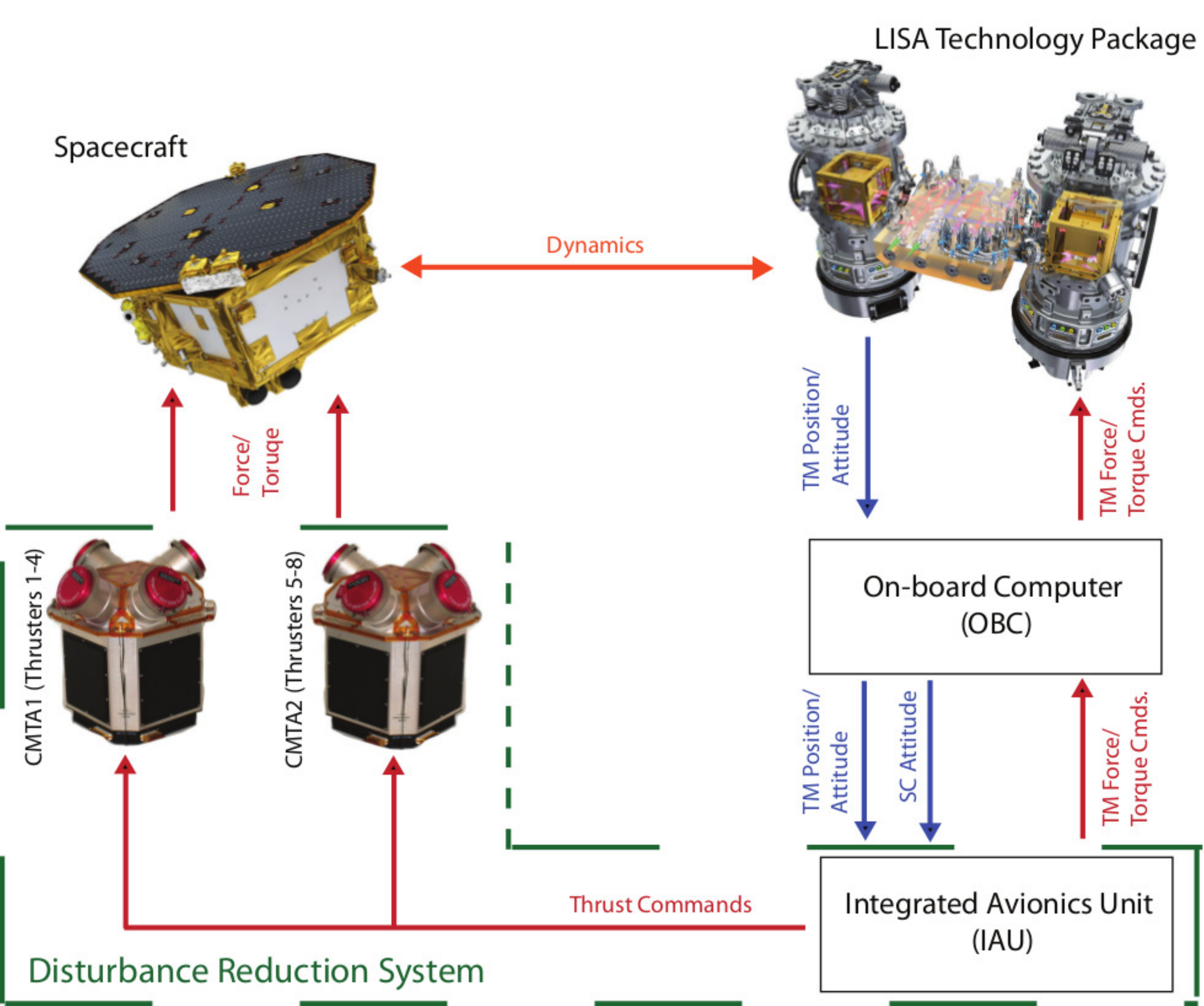}
\caption{Block diagram of the Disturbance Reduction System (DRS) and its interfaces with the LISA Pathfinder spacecraft and the LISA Technology Package instrument. Renderings of LPF and LTP courtesy of ESA/Medialab.}
\label{fig:block_diagram}
\label{default}
\end{center}
\end{figure}

Figure~\ref{fig:block_diagram} shows a block diagram of the DRS hardware and its major functional interfaces to the LPF spacecraft and the LTP instrument. The DRS consists of three distinct hardware units: the IAU and two Colloidal Micronewton Thruster Assemblies (CMTAs) with four thrusters each.  The IAU interfaces with the primary LPF computer, known as the On-Board Computer (OBC), and the OBC provides interfaces to the LTP instrument as well as other spacecraft systems such as the star tracker and communications systems. In drag-free operations when the DRS is in control of the spacecraft attitude, the LTP provides measurements of the position and attitude of the two test masses, which are processed by the OBC and sent to the IAU along with spacecraft attitude measurements derived from the LPF star trackers. This information is processed by the Dynamic Control System (DCS) software running on the IAU, which determines the appropriate forces and torques to apply to the spacecraft and the test masses. Test mass force/torque commands are sent by the IAU to the OBC, which relays them to the GRS front-end electronics within the LTP.  Spacecraft force/torque commands are decomposed into individual CMNT thrust commands, which are then sent to the CMNTs. \footnote{After the anomaly experienced by CMTA2, the commands sent by the IAU to the CMTAs were actually low-level current and voltage commands that are functionally equivalent to thrust commands. See Sec.~\ref{sec:ANOMALIES} on anomalies and recovery for more detail.}

The DRS is a single-string system, but with a redundant RS-422 communication interface between the IAU and the OBC, and redundant IAU DC/DC power converters and thruster power switches.
The  redundant power busses are cross-strapped to each thruster cluster, which are single string. The A-Side power bus was the primary bus used during the mission.

\section{The Colloidal Micro-Newton Thruster Assemblies}
\label{sec:CMNT}
\subsection{Components}
Colloid thrusters were selected to be developed by ST7-DRS because of their potential for extremely high precision thrust; extremely low noise; and a larger specific impulse compared to cold-gas systems ($\sim240\,$s vs $\sim70\,$s). Colloid thrusters are a type of electrospray propulsion, which operate by applying a high electric potential difference to charged liquid at the end of a hollow needle in such a way that a stream of tiny, charged droplets is emitted generating thrust. An advantage of this system is that the liquid colloidal propellant can be handled with a compact and lightweight propellent management system and requires no pressure vessels or high temperatures. The requirement for high-voltage power supplies is a disadvantage. Colloidal thrusters can be designed to operate in various thrust ranges according to the number of needles that are used in each thruster head. The ST7-DRS configuration, developed specifically for ST7s performance requirements by Busek, provides a thrust range  from 5 to 30 $\mu$N per thruster (larger thrusts are achievable in diagnostic mode).

\begin{figure}
\begin{center}
\includegraphics[width=\columnwidth]{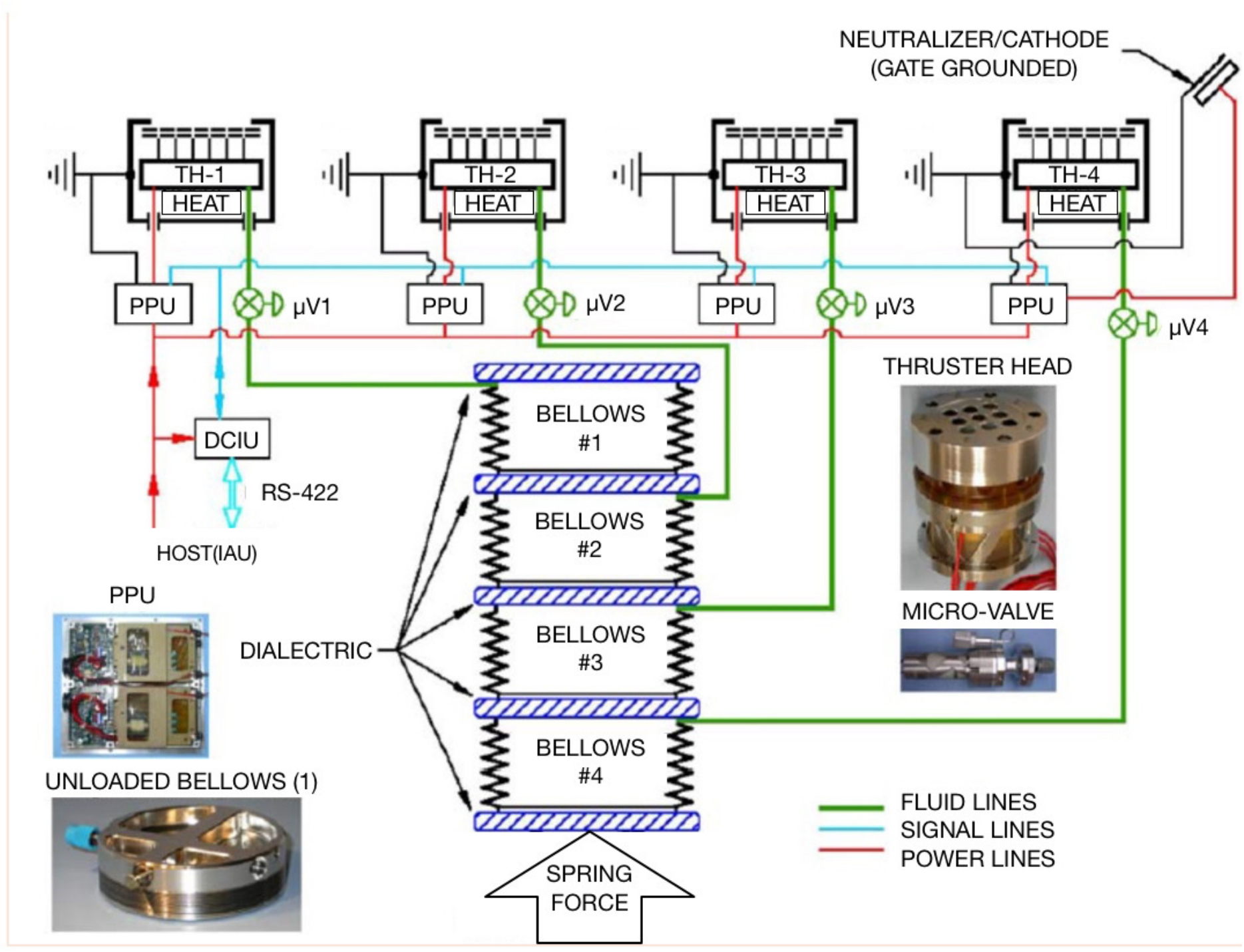}
\caption{CMNT propulsion system components and configuration. The Carbon nanotube cathode is not shown.
 \label{fig:cmnts} }
\label{default}
\end{center}
\end{figure}

DRS includes two Colloidal Micro-Newton Thruster Assemblies (CMTAs), each of which includes: 4 thruster heads, 4 propellant feed systems, 4 Power Processing Units (PPUs), 1 cathode, and 1 Digital Control Interface Unit (DCIU)~\cite{Ziemer2008}. Figure \ref{fig:cmnts} shows both a  block diagram for one of the 4-thruster assemblies. Each thruster head includes a manifold that feeds nine emitters in parallel, a heater to control propellant temperature and physical properties, and electrodes that extract and accelerate the propellant as charged droplets. The thruster heads are fed by independent bellows and microvalves (the feed system).  The propellant is the room temperature ionic liquid 1-ethyl-3methylimidazolium bis(triflouromethylsulfonyl)imide (EMI-Im)
and is stored in four electrically isolated, stainless steel bellows, which use compressed constant-force springs to supply the four microvalves with propellant at approximately 1 atm of pressure. The propellant flow rate is controlled by a piezo-actuated microvalve. The thruster heads and feed system voltages are independently controlled through the PPUs, which are controlled, in turn by the DCIU. The DCIU has an on-board PROM (programmable read-only memory) that stores the thruster operating software and control algorithms. The DCIU has power, command, and telemetry interfaces to the IAU. The CMTA mass is 14.8 kg with $\sim0.5\,$kg of propellant distributed into each of the 4 thruster bellows. The nominal power consumptions of each CMTA is $\sim17\,$W. 

Each CMTA also includes one propellantless field emission cathode neutralizer, included to neutralize the emitted spray of charged droplets after they are accelerated, and so prevent spacecraft charging by the thrusters. The cathode neutralizers are fabricated from a carbon nanotube (CNT) base with an opposing gate electrode controlled by the DCIU. Each CNT is capable of producing 10 $\mu$A to 1 mA  using extraction voltages of 250 to 800 V. The neutralizer was tested during the extended mission and produced the desired current. As expected, the measured spacecraft charging with respect to the test masses \cite{LPF_CHARGE_PRD_2018} indicated that the induced spacecraft charge rate was larger in magnitude than and opposite in sign from the effect of the CMNTs, meaning the neutralizer was not necessary for maintaining spacecraft charge control. 
 
\subsection{Thrust model}

The thrust (T)  from each CMNT is approximated by~\cite{ziemer2009performance, demmons2008st7}:

\be\label{eq:T=CIV}	
T =C_1\,  I_B^{1.5} \, V_B^{0.5} \, ,			
\ee
where $I_B$ is the total beam current from the $9$ emitters, $V_B$ is beam voltage, and $C_1$ is the thrust coefficient. $C_1$ depends mostly on physical properties of the propellant 
(viscosity, electrical conductivity, etc.) and also on the characteristics of the plume (beam divergence, charge-to-mass ratio distribution, etc.). In operation, the DCIU adjusts the the beam voltage (2-10 kV) and propellant flow rate for each thruster head to achieve the desired thrusts. The mass flow rate is not measured directly; instead the beam current $I_B$ is measured and controlled by actuating a piezo microvalve. $I_B$ is controlled to better than 1 nA  over the operating range of 2.25 to 5.4$\mu$A, corresponding to a thrust resolution of  $\le 0.01\,\mu$N. Independent, fine control of both the beam voltage and beam current allow for precise control of thrust to better than 0.1 $\mu$N,  with $<$0.1 $\mu\textrm{N}/\sqrt{\textrm{Hz}}$ thrust noise.

The value $C_1$ is temperature-dependent; $C_1$ decreases with increasing temperature. Thrust-stand measurements on Engineering Model (EM) units validated this model to a precision of $\sim\,$2\%, consistent with the calibration of the thrust stand\cite{demmons2008st7}. The best-fit value for $C_1$ under nominal beam voltage and current and with a  propellant temperature $T_{P} = 25^{\circ}$C was $31.9\,\textrm{N}\textrm{A}^{-3/2}\textrm{V}^{-1/2}$.  A substantial portion of the DRS operations was utilized to perform calibration experiments to validate this model in flight as well as explore potential higher-fidelity models. These experiments and their results are discussed in section \ref{sec:thustCal}.

\section{The Dynamic Control System (DCS)}
\label{sec:DCS}
The DRS control system maintains the attitude of the spacecraft, as well as the position of the test masses within their housings, both by moving the spacecraft via the CMNTs and by moving the test masses via electrostatic actuation. There were initially six DRS Mission Modes managed by the IAU: Standby, Attitude Control, Zero-G, Drag-Free Low Force, 18-DOF (Degree of Freedom) Transitional, and 18-DOF Mode. One additional mode,  the Zero-G Low Force was added for the extended DRS mission, to provide improved performance in the accelerometer mode, where some of the most sensitive thruster experiments were performed. 

Standby mode is used when IAU is powered on but no actuation commands are being generated by the control system, generally when the LTP controller is active.  Attitude Control Mode is used for the transition from LTP control to DRS Control.  In this mode, the DRS nulls spacecraft attitude errors and their rates using the thrusters, while electrostatically suspending the both test masses. In the Zero-G Mode, disturbance forces on the spacecraft, such as those from solar radiation pressure, are nulled out in a low-bandwidth loop that minimizes common-mode actuation on the test masses by applying forces on the spacecraft using the CMNTs. The Zero-G low force variant utilized the same control scheme but with the GRS actuation set to its high resolution / low force authority setting. In the Drag Free Low Force (DFLF) Mode the spacecraft's position is controlled via the CMNTs to follow the reference test mass (RTM, configurable to be either of the two LTP test masses) in all translational axes. Hence, it is the lowest mode in which drag-free flight of a single test mass is achieved. The 18-DOF Transitional is, as its name suggests, a transitional mode to get from DFLF to 18-DOF control of the spacecraft and the two test masses. In the 18-DOF mode, the DRS uses the thrusters to force the spacecraft to follow the RTM, i.e., to maintain the nominal gap of the RTM with respect to its housing along all three axes. The DRS uses the torque from the CMNTs to control the spacecraft attitude,  in the measurement band (1-30mHz), so that it follows: (a) the non-reference test mass (NTM) in the transverse directions (normal to the LTP axis); and (b) the relative attitude of the RTM about the sensitive axis. The orientations of both test masses are controlled via electrostatic suspension below the measurement bandwidth (MBW).  Further details on the spacecraft and test mass control design for each mode may be found in \cite{Maghami2004,Hsu2004}. 

The DRS baseline architecture made use of only the capacitive sensing measurements of the test mass positions and orientations, which was the configuration for both ST7 and LTP when the DRS design was consolidated.  After successfully commissioning DRS in this configuration, the system was modified to use the higher-precision interferometric data from LTP for the degrees of freedom where it was available. 

\section{Characterization of Thruster Performance}
\label{sec:CMNTP}
In-flight characterization of the CMNT technology was a major goal of the ST7 mission. This section summarizes the experiments conducted during DRS operations and the top-level results. The CMNT properties investigated during the flight campaign include thrust range, response time, calibration, and thrust noise.  In general, two sources of data were available for these investigations. The first was internal thruster telemetry such as beam currents, beam voltages, valve voltages, temperatures, etc. These quantities can be used to estimate the CMNT thrust using physics-based models which were validated during the CMNT development with thrust stand measurements. The second source of data, which is unique to this flight test, was the rest of the Pathfinder spacecraft, in particular the test masses and interferometer of the LTP.  Measurements with LPF and LTP data allowed the CMNT thrust model to be independently validated and its exquisite sensitivity allowed thrust noise measurements in the LISA band at a level never achieved with ground-based thrust stands.  

\subsection{Functional Tests: Range and Response Time}
\label{sec:CMNTfunc}
\begin{figure}
\begin{center}
\includegraphics[width=\columnwidth]{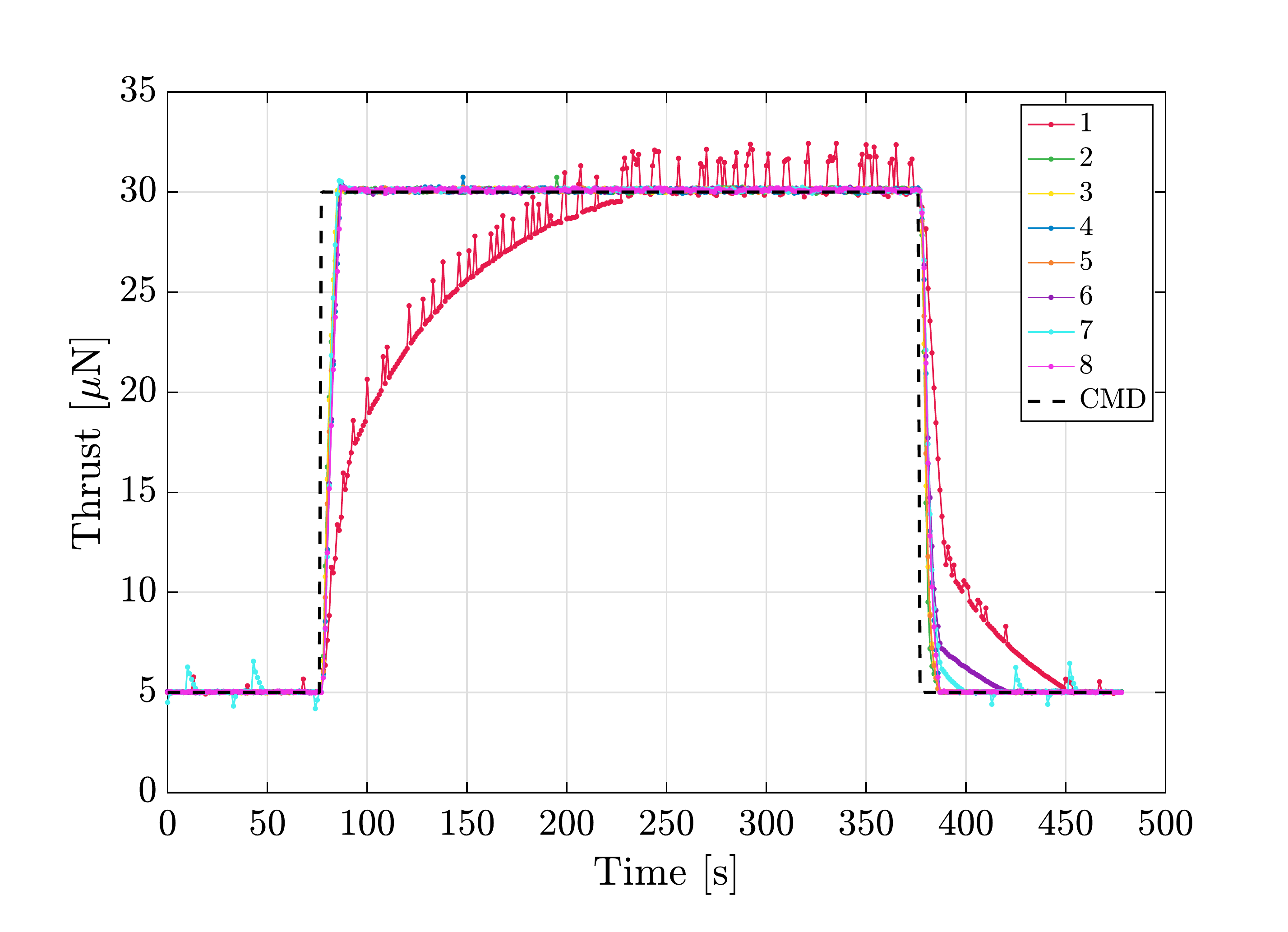}
\caption{Full-range response test for all eight CMNTs conducted as part of initial thruster commissioning in January 2016 prior to separation of the LPF Propulsion Module. All 8 thrusters demonstrated the full thrust range, although CMNT\#1 had an abnormally slow response time, perhaps due to some obstruction. In addition CMNT\#1 exhibited a `blipping' mode consistent with one of the nine emitter tips cycling between a spraying and non-spraying state.  \label{fig:cmnt_resp} }
\label{default}
\end{center}
\end{figure}

As described in Table \ref{tab:dates}, the CMNTs were commissioned in two phases in 2016.  The first phase, in January 2016, was conducted prior to separation of the propulsion module so that the CMNTs would be available to serve as a backup propulsion system for LPF, should the primary cold-gas system experience an anomaly after separation.  Figure \ref{fig:cmnt_resp} shows a full-range response test  in which all 8 CMNTs are initially at their minimum thrust level of 5$\,\mu$N and then commanded to their maximum level of 30$\,\mu$N for 300$\,$s before returning to 5$\,\mu$N.  The thrust command (same for all thrusters) is shown in black and the predicted thrust based on CMNT telemetry and the ground-validated model is shown in colored lines. This initial test demonstrated that all eight CMNTs could operate in their requisite 5-30$\,\mu$N  thrust range. With the propulsion module still attached to the spacecraft and the LTP instrument not yet commissioned, there was limited availability of precision data from the platform. However, telemetry from the propulsion module ACS thrusters and from the LPF star tracker showed force/torque motions on the platform that were roughly consistent with thrust levels commanded to the CMNTs.\\

Both the thruster control law and the dynamics of the thruster head are expected to lead to a delayed response time, for which the design requirement was 10$\,$s. As shown in Figure \ref{fig:cmnt_resp}, 7 of the 8 CMNTs meet this requirement for the response time from 5$\,\mu$N to 30$\,\mu$N. The exception is CMNT\#1, which has a response time of $\sim 170\,$s.  As discussed in Appendix A.1,  this delayed response is consistent with some obstruction in the CMNT\#1 feed system.  The response time for 30$\,\mu$N to 5$\,\mu$N was slightly longer for all CMNTs and significantly longer for thrusters 1,6, and 7. This is likely due to the response of the piezo microvalve, which is controlled by a PID loop to maintain the desired current level. The valve actuator is encased in a potting compound that provides electrical isolation but also adds some mechanical compliance to the valve. In addition, there is some variation in the pre-load and piezo response from valve to valve that results in different flow response as a function of valve voltage. An examination of the valve voltage during this experiment reveals that while the voltage for valves 1,6, and 7 dropped to zero after the transition from commands of 30$\,\mu$N to 5$\,\mu$N, the electrometer still measured a slowly-declining current after the valve voltage reached zero.  This suggests that the piezo actuators in these valves were unable to close the valves to the desired position until the valve mechanically relaxed, after which time the piezo could begin to actuate again. \\
Finally, CMNT\#1 also shows an impulsive behavior known as `blipping', which is caused when the number of emitters actively flowing propellant in the CMNT head changes.  In standard operations, all 9 emitters should be expelling propellant for the full range of thrusts. However, if one emitter has a significantly increased hydraulic resistance due to an obstruction, it can periodically stop and start, leading to an abrupt change in the thrust. While efforts were made during the mission to improve the performance of CMNT\#1 during the mission, neither the response time not the rate of `blipping' significantly improved. This is further discussed in Appendix A.1.   Note that there is also some evidence of `blipping' in in CMNT\#7 in Figure \ref{fig:cmnt_resp}, although at a much lower rate than for CMNT\#1 and also only at the minimum thrust level.  The blipping behavior for CMNT\#7 rapidly improved as commissioning proceeded and was not observed during the remainder of the mission, suggesting that the (presumably much smaller) obstruction that was responsible was cleared.  

\begin{figure}
\begin{center}
\includegraphics[width=\columnwidth]{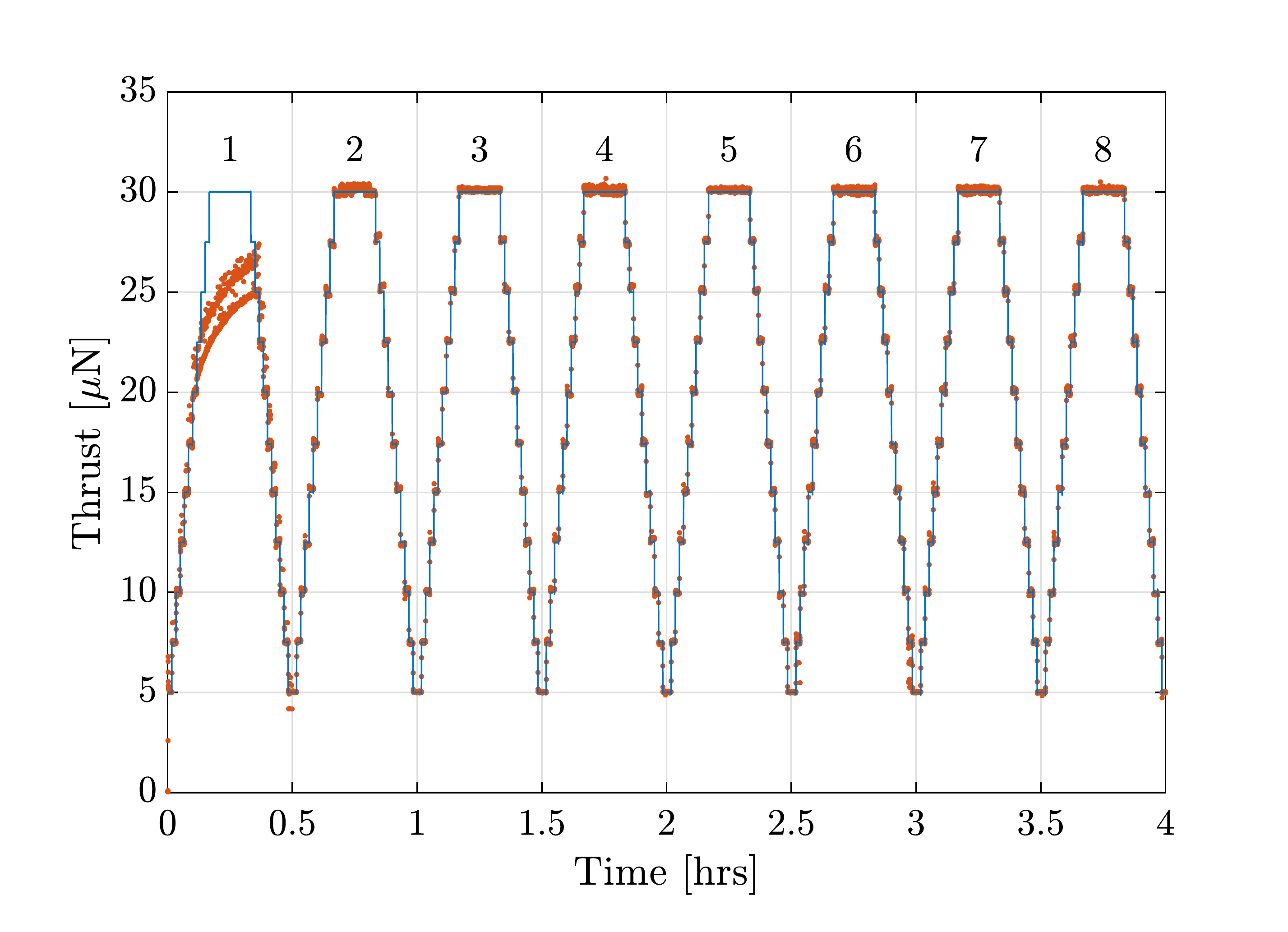}
\caption{Thruster actuation test conducted during DRS commissioning in July 2016. Each of the eight CMNTs was successively cycled from 5$\,\mu$N to 30$\,\mu$N and back in 5$\,\mu$N steps. The achieved thrust estimated from the CMNT telemetry closely matches the thrust commands with the exception of CMNT\#1, which exhibits both a slow response time and 'blipping' consistent with one of its nine emitters firing only intermittently. \label{fig:cmnt_range} }
\label{default}
\end{center}
\end{figure}

After the propulsion module was successfully separated and the spacecraft was under control of the cold-gas micropropulsion, the CMNTs were placed in a safe mode for approximately 6 months of LTP operations. In July 2016, the second phase of CMNT commissioning was conducted to prepare for DRS operations. Figure \ref{fig:cmnt_range} shows a thruster functional test in which each CMNT is successively ramped, in 5$\,\mu$N increments, from 5$\,\mu$N to 30$\,\mu$N and back.  The blue line shows the thrust command and the red data shows the estimated thrust  based on the CMNT telemetry and the ground-validated model. Again, 7 of the 8 CMNTs perform as designed, but CMNT\#1 exhibits both the episodic `blipping' and an overall slow response time. We note that here the 30$\,\mu$N limit on the maximum thrust was set by the flight software. In the extended mission, the diagnostic mode of the CMNTs was used to manually command the current and voltage to demonstrate extended thrust range. CMNT\#2 and CMNT\#5 were stepped up to 40$\,\mu$N early in the extended mission and then CMNT\#5 was stepped up to 50$\,\mu$N and CMNT\#2  was stepped up to 60$\,\mu$N near the end of the extended mission. The CMNTs passed all of these extended-range tests without incident.

\subsection{Thruster Calibration Measurements}
\label{sec:thustCal}
\vskip 0.2in
As mentioned above, the additional instrumentation on the Pathfinder spacecraft, in particular the LTP, allows use of the spacecraft as a low-noise thrust stand that can be used to calibrate the thrust model.  This section describes the design, analysis, and results of thruster calibration experiments carried out using this method.

\subsubsection{Experiment design}
All the calibration experiments had the following general form.  The thrust command to one of the eight CMNTs is modulated by some sinusoidal or square wave, which we refer to as the \emph{injection waveform}, and each thruster's calibration constant is derived from the resulting modulations on i) the motion of the spacecraft  , ii) the motion of the two TMs, and iii) the electrostatic forces on the TMs.  The full set of injection waveforms that we used is summarized in Table \ref{tab:T1}.  In each experiment, we cycle through each of the thrusters one at a time.

We chose to perform  the majority of the injections in accelerometer control mode, out of concern that drag-free control would lead to a complicated mixing of the injection signals across all thrusters simultaneously, as well as suppressing the modulation on the main thruster of interest.  Operating in accelerometer mode is very close to the standard operating mode in the ground-based thrust stand measurements.

The amplitude and frequency of the injections were selected to balance the needs of the system (low disturbance, slew rate limits, and available experiment time) against the needs of the analysis, as characterized by the expected signal to noise ratio (SNR). Our baseline approach was to perform the experiment in the DRS's high-force actuation mode, since it provides higher control authority and therefore should permit larger-amplitude injections without losing stability. Later in the mission, a set of injections was designed for the DRS's low-force actuation mode, which had a better-characterized calibration of the applied test mass forces and torques than the high-force mode.  

Early in DRS operations, each injection set was demonstrated on a subset of thrusters to assess the quality of the response and resulting analysis. Set 1, with the most gentle system response but longest duration, was used for initial checkout and sets 2 through 4 were used together to more rapidly characterize thruster performance. Set 5 utilized a square wave to measure the response at multiple Fourier frequencies and was utilized in some limited tests. In this paper, we present results from the `standard' suite of sets 2 through 4, which were used for the majority of the investigations in both baseline and extended DRS operations. 

\begin{table}
\caption{Waveforms of thruster calibration experiments.\label{tab:T1}}
\begin{center}
\begin{tabular}{|c|c|c|c|c|c|c|}
 \hline
Set \#  & Waveform &   Frequency &   Amplitude  &   Duration & $\rho_{HF}$ & $\rho_{LF}$  \\ \hline
1 & sine & 23 mHz & 1 $\mu$ N & 5220 s & 200 & 500 \\ \hline
2 & sine  & 23 mHz & 3 $\mu$ N &  696 s  & 200 & 500 \\ \hline
3 & sine & 29 mHz  & 5 $\mu$ N &  552 s  & 200 & 800 \\ \hline
4 & sine  & 40 mHz  & 3 $\mu$ N &  600 s & 66 & 400 \\ \hline
5 & square &  23 mHz & 3 $\mu$ N & 696 s & ~200 & ~500 \\ \hline
 \end{tabular}
\end{center}
\label{default}
\end{table}

\subsubsection{Calibration results}

The basic analysis approach is to estimate the acceleration of the LISA Pathfinder spacecraft using the LTP data, and compare that with the thrust derived from the thruster's measured $V_B$ and $I_B$ and the thrust model, Eq.~(\ref{eq:T=CIV}). As shown in Appendix B, using the average acceleration of the two test masses causes most of the rotating-frame effects to cancel out, making the results more robust against systematic errors.

\begin{figure}
\begin{center}
\includegraphics[width=\columnwidth]{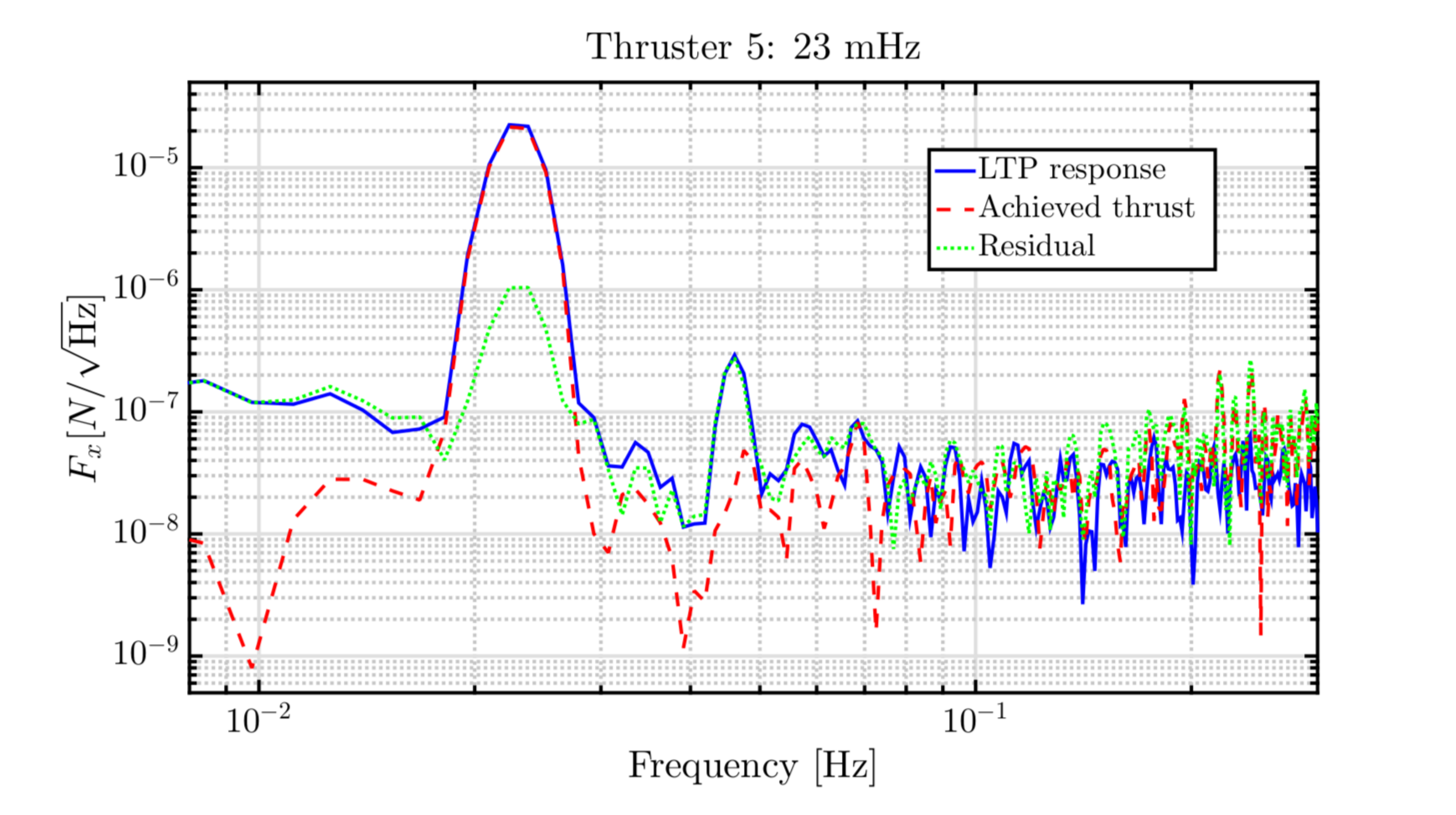}
\caption{Thrust spectrum as measured from LTP data (blue), thrust based on thrust model and best-fit C1 value(red), and residual (difference) between those two (green), for a 23 mHz injection in \emph{high-force} actuation mode in CMNT\#5. \label{fig:t1psd} }
\label{default}
\end{center} 
\end{figure}

\begin{figure}
\begin{center}
\includegraphics[width=\columnwidth]{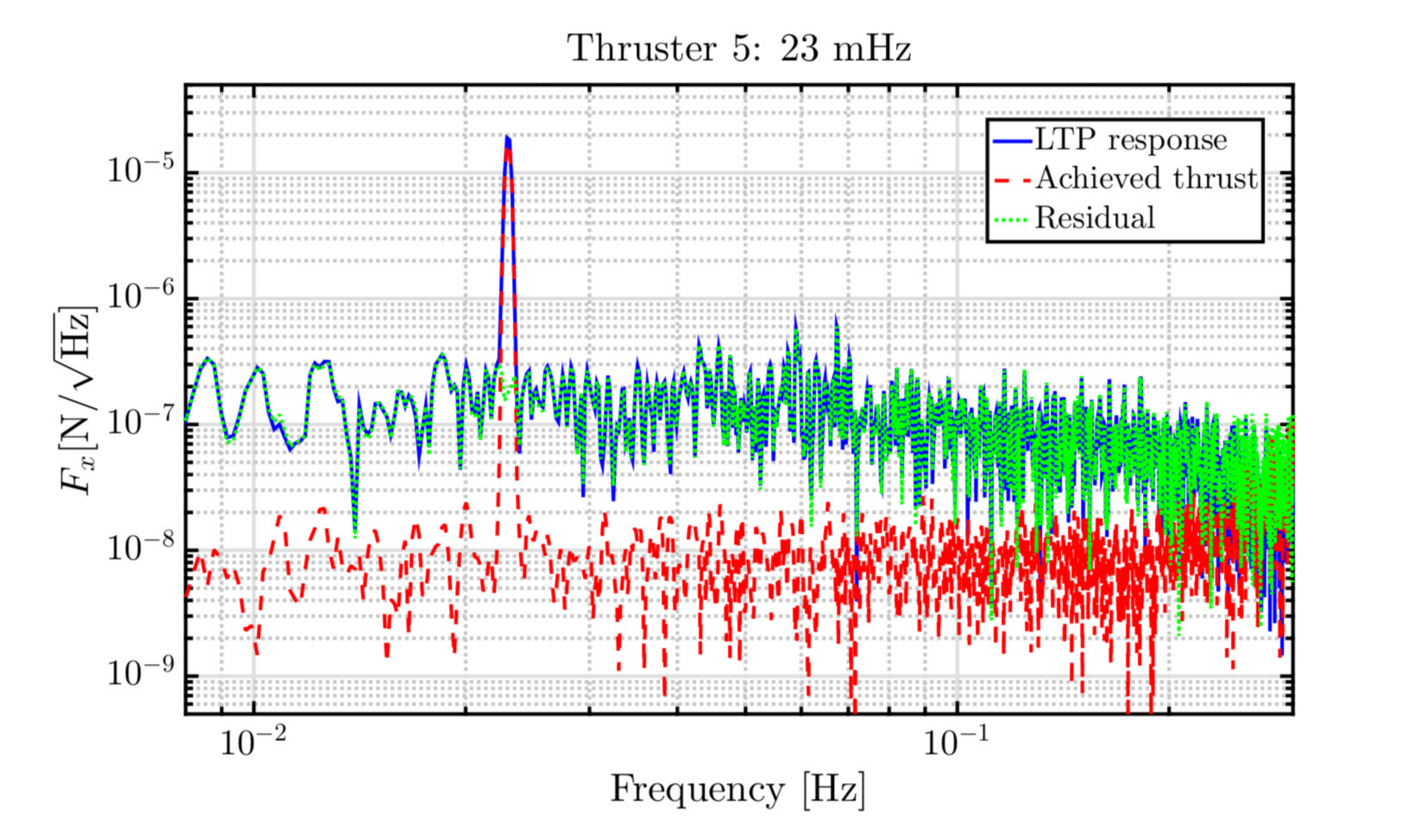}
\caption{Thrust spectrum as measured from LTP data (blue), thrust based on thrust model and best-fit C1 value(red), and residual (difference) between those two (green), for a 23 mHz injection in \emph{low-force} actuation mode in CMNT\#5.  \label{fig:EM8psd} }
\label{default}
\end{center}
\end{figure}

We use a Markov-Chain Monte-Carlo method to estimate the maximum-likelihood gain and delay of each thruster, jointly fit across the three injection frequencies. Figure \ref{fig:t1psd} shows an example fit for CMNT\#5 for an injection with the TMs in the high-force mode. The injection signal is suppressed by a factor of roughly 20, although it is still visible in the residual. Figure \ref{fig:EM8psd} shows a similar fit for CMNT\#5 for an injection with the TMs in the low-force mode.  Here the injection signal is suppressed by a factor of roughly 100, and no residual is visible.

 Table \ref{tab:C1} presents the best-fit value for the C1 coefficient in Eq.~(\ref{eq:T=CIV}) for each thruster. These values are averaged over 4 measurements, except for CMNT\#4 which was only averaged 3 times due to the propellant bridge (see Appendix A.3 for a discussion of this anomaly).  All of the results are in the range $29\sim32\,\textrm{N}\textrm{A}^{-3/2}\textrm{V}^{-1/2}$, roughly consistent with the value of $31.9\,\textrm{N}\textrm{A}^{-3/2}\textrm{V}^{-1/2}$ derived from the ground tests \cite{Ziemer2010}.

There is also a small but statistically-significant discrepancy between the experiments conducted in high-force and low-force modes, which suggests that there is some aspect of the calibration that has not been properly accounted for.  Note that there is no low -force mode measurement for CMNT\#4, as this experiment was implemented after it was disabled.  We do not report the best-fit delays, since in addition to physical delays, they include relative delays between the DRS data packets containing the CMNT telemetry and the LTP data packets containing the LTP telemetry, and the latter are not physically relevant.  
 
\begin{table}
\caption{Summary of thruster calibration results. C1 values are in units of [$ N A^{-3/2} V^{-1/2}$].\label{tab:C1}}
\begin{center}
\begin{tabular}{|c|c|c|c|c|}
\hline
T\# & T1 C1 & error & T1 (LF) C1 &  error  \\
\hline
1 & 31.52 & 0.15 & 31.91 & 0.10 \\ \hline
2 & 30.18 & 0.10 & 30.97 & 0.10 \\ \hline
3 & 28.78 & 0.09 & 32.12 & 0.10 \\ \hline
4 & 29.92 & 0.09 & - & - \\ \hline
5 & 29.96 & 0.10 & 31.49  &  0.10 \\ \hline
6 & 30.02 & 0.10 & 30.90 &  0.10 \\ \hline
7 & 29.71 & 0.09 & 30.37 &  0.10 \\ \hline
8 & 29.86 & 0.10 & 30.53 & 0.10 \\ \hline 
\hline
\end{tabular}
\end{center}
\label{default}
\end{table}

\subsubsection{Temperature Dependence}
\label{sec:CMNT_T}

\begin{figure}
\begin{center}
\includegraphics[width=\columnwidth]{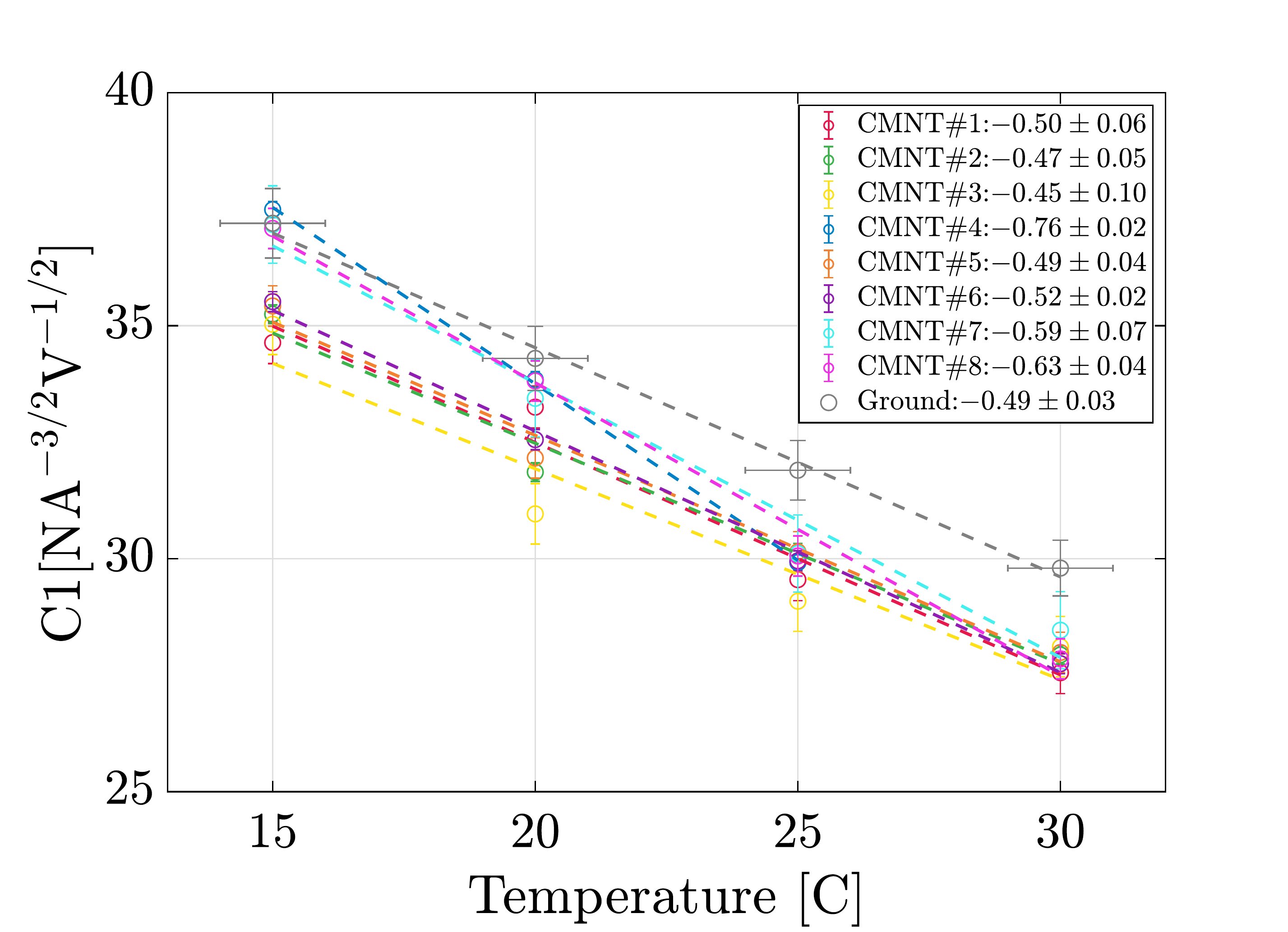}
\caption{ Comparison of measured and expected dependence of thruster coefficient on temperature. Legend entries show the best-fit linear coefficient for each CMNT and the ground test in units of $N A^{-3/2} V^{-1/2} C^{-1}$. \label{fig:C1temp} }
\label{default}
\end{center}
\end{figure}

The $C1$ coefficient in the CMNT thrust model, Eq.~(\ref{eq:T=CIV}), is expected to depend on the propellant temperature, which is set to $25\,$C during most of the mission.  To validate this model, a campaign was undertaken to alter the temperature set point using onboard heaters, and measure the CMNT calibration using injection sets 2 through 4. Figure \ref{fig:C1temp} shows the measured calibrations at temperatures of $15\,$C, $20\,$C, $25\,$C, and $30\,$C along with a linear fit for temperature dependance. Similar data obtained on the ground with thrust-stand measurements is included for comparison. Note that due to the CMNT\#4 propellant bridge, there are no values for $30\,$C for that thruster. Additionally, several other measurements were made in the initial calibration experiment at the nominal temperature of $25\,$C, which are included for all thrusters.

\subsection{Thruster noise performance}
\label{sec:CMNTnoise}
The DRS had a Level 1 performance requirement to demonstrate a spacecraft propulsion system with noise less than 0.1$\, \mu\textrm{N}/\sqrt{\textrm{Hz}}$ over a frequency range of 1 mHz to 30 mHz.   We use two different approaches to estimate the CMNT noise performance.  The first method uses the CMNT flight data and the thrust  model in Eq.~\ref{eq:T=CIV}, using the calibration results for the $C1$ coefficients.  To estimate the intrinsic thruster noise apart from the required spacecraft control, we subtract the thrust commands.  This provides the thrust error. This represents a lower limit on the thrust noise, since additional effects not captured by the measured $I_B$ and $V_B$ values could produce additional noise. Some such effects, such as CMNT shot noise, are known to be well below the measured noise floor, but there is also the possibility of unmodeled noise. 

Our second approach for estimating the CMNT noise is to use the Pathfinder spacecraft as a thrust stand, as was done for the thruster calibration measurements.  In the measurement band, $1-30$mHz, thrust noise is expected to dominate the total budget of force noises on the spacecraft  .  Measuring the acceleration noise of the spacecraft should therefore be an effective way to estimate the thruster noise, and is formally an upper limit.  Unfortunately, this approach is complicated by the fact that after the anomaly experienced by CMNT\#4 (see Appendix A), some portion of the cold-gas micropropulsion system was required to be active whenever the CMNT system was active. The thruster noise measured during these periods includes contributions from the cold-gas system as well as the CMNTs.

\subsubsection{CMNT Noise from internal telemetry}
\label{sec:IVnoise}

\begin{figure}
\begin{center}
\includegraphics[width=\columnwidth]{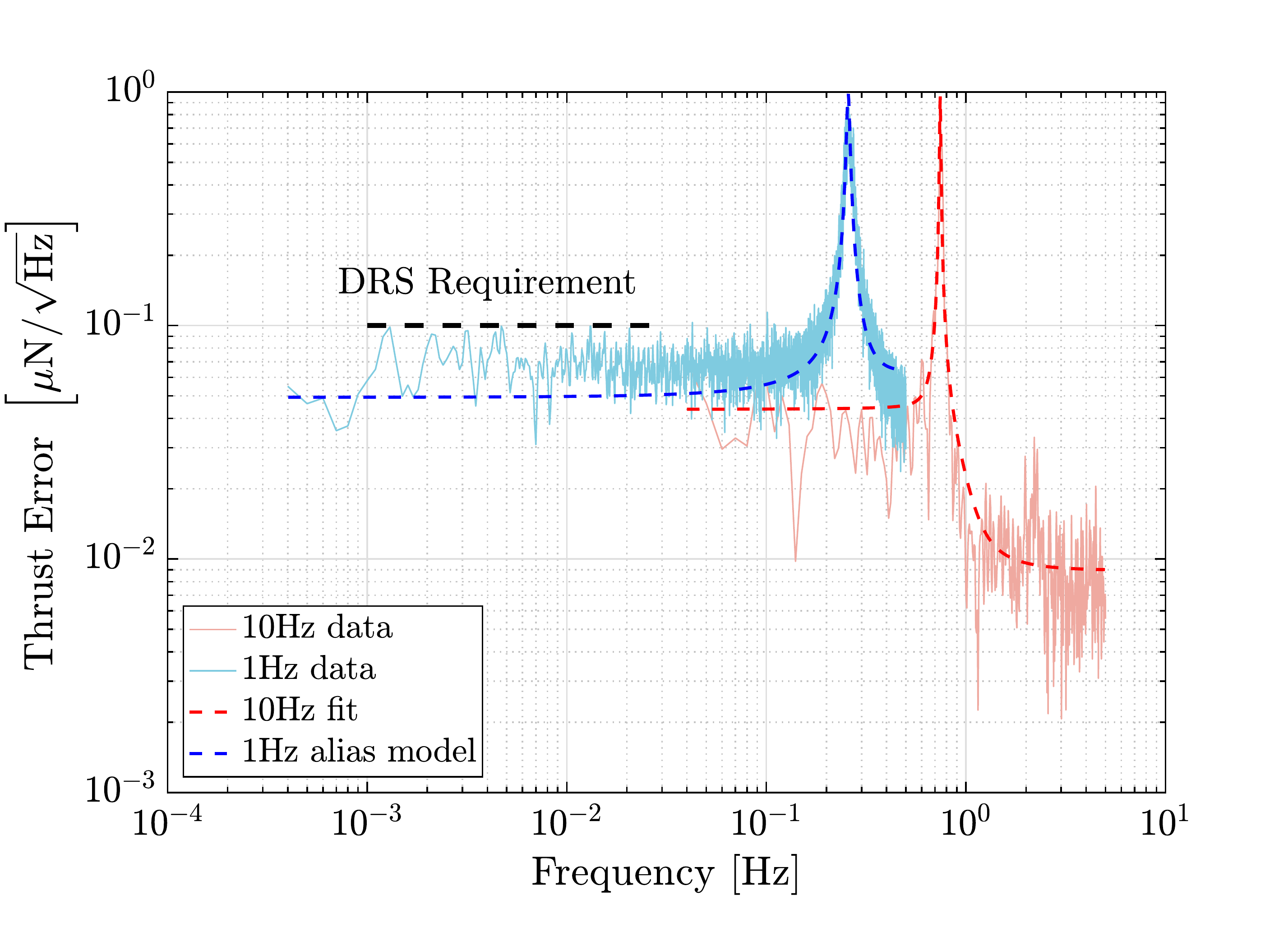}
\caption{Measured thrust error (thrust command - modeled thrust) in CMNT\#5 for an 8-hour period on 2017-04-24. The light blue trace is the full duration, sampled at 1Hz. The light red trace is for a adjacent 200s segment sampled at 10Hz. The dashed red line is a fit to the 10Hz spectrum and the dashed blue is a model of how that spectrum gets aliased by downsampling the data without using an anti-aliasing filter.}
 \label{fig:CMNT5_alias} 
\end{center}
\end{figure}

The light blue trace in Figure \ref{fig:CMNT5_alias} shows the measured amplitude spectral density of the thrust noise for CMNT\#5, using the CMNT flight data estimation.  This data, sampled at 1 Hz, comes from an 8-hour period on 2017-04-24, while in the 18DOF controller configuration. Below 100$\,$mHz, it is a flat with a level of approximately 70$\,\textrm{nN}/\sqrt{\textrm{Hz}}$.  Note that this individual thrust noise is somewhat better than the requirement. Additionally, around 250$\,$mHz, well above the measurement band, the thrust error for CMNT\#5 exhibits a sharp spectral line feature. The light red trace in Figure \ref{fig:CMNT5_alias} shows the thrust error spectrum from a 200s period of 10Hz-sampled data taken just after the data for the blue trace.  In the red trace, the line is shifted to 750$\,$mHz and the flat level at lower frequencies is reduced.  This already strongly suggests that the 250$\,$mHz in the light-blue trace is actually a 750$\,$mHz effect that is getting aliased down to  250$\,$mHz in the 1Hz-sampled data.   The CMNT telemetry, which is delivered to the IAU at 10Hz, is typically decimated to 1Hz without the use of an anti-aliasing filter.  The rationale for this decision was that the CMNT thrust model depends non-linearly on the beam voltage and current and any averaging or other filtering operation applied to the current and voltage would not give the correct result for the thrust model. In addition, neither the current nor voltage telemetry was expected to have significant power above 0.5Hz. To confirm this interpretation, we fit a spectral model to the 10Hz data (dashed red line in Figure \ref{fig:CMNT5_alias} and compute the aliased version of that spectrum, shown by the dashed blue line in Figure \ref{fig:CMNT5_alias}).  Clearly, this reproduces the 250$\,$mHz peak seen in the 1Hz data.  Based on this analysis, we estimate that the intrinsic noise of CMNT\#5 is closer to 40$\,\textrm{nN}/\sqrt{\textrm{Hz}}$ in the absence of aliasing. It is important to note that aliasing only affects the \emph{telemetered} values of beam current and voltage. The on-board processing is done at the full $10\,$Hz rate.

\begin{figure}
\begin{center}
\includegraphics[width=\columnwidth]{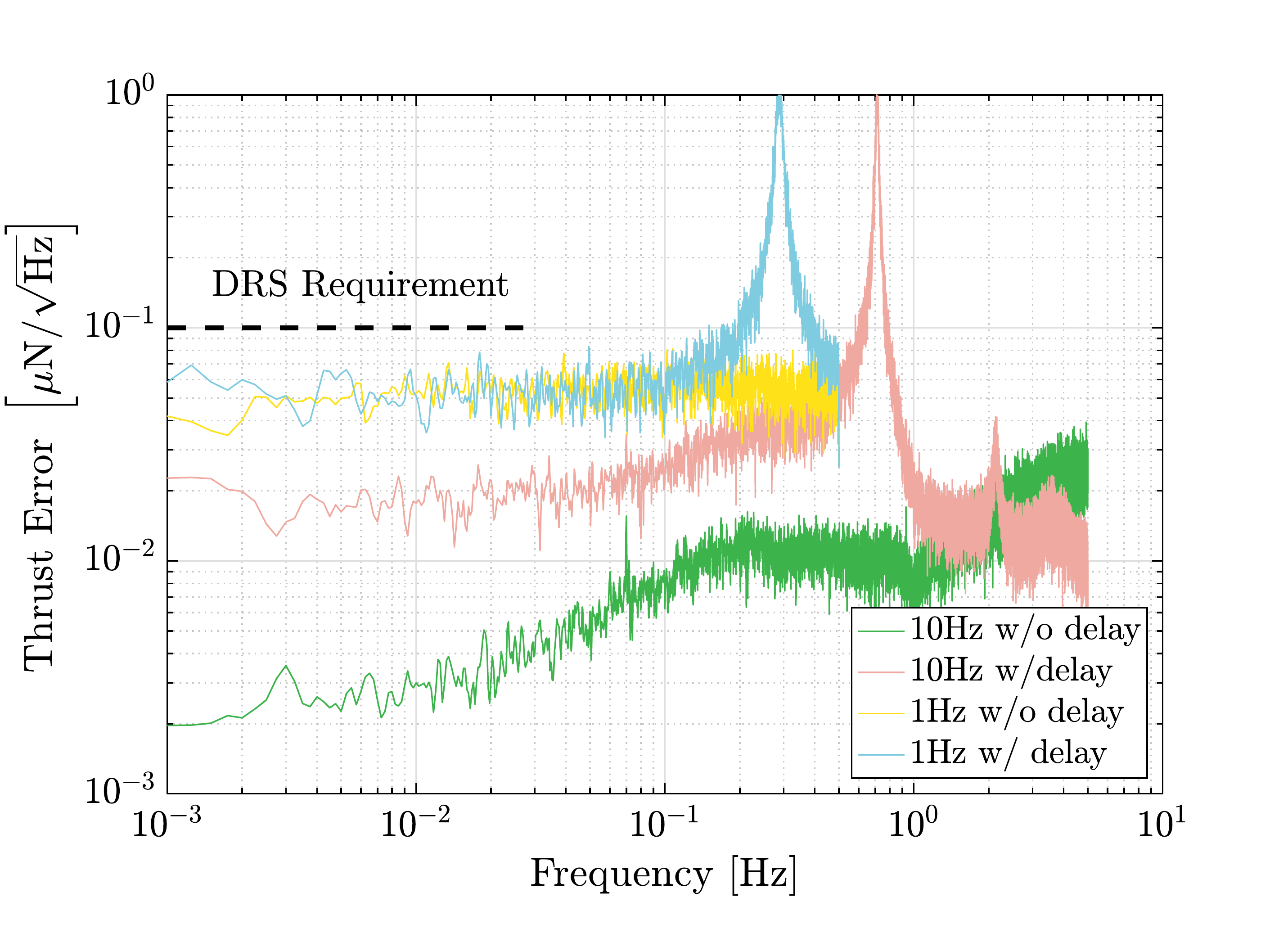}
\caption{Simulated thrust error (thrust command - modeled thrust) for CMNT\#5 thrust commands using a software model of the thrust control algorithm. The green trace shows the expected thrust error in the baseline case of minimal delay between the commands and response of beam voltage and current. The light red trace shows the same signal when a 300ms delay is introduced in the beam voltage and current response. The blue and yellow traces show the with- and without-delay signals, respectively, downsampled to 1Hz without anti-aliasing filters}
 \label{fig:CMNTsim} 
\end{center}
\end{figure}

The source of the peak in the 10Hz data, which was not observed in ground testing, is suspected to be an oscillation in the thruster control loop which arose when the thrust control algorithm was moved from the DCIU to the IAU, following the DCIU anomaly described in Appendix A. This switch resulted in an additional delays--first for the current and voltage commands to travel from the IAU to the DCIU, and then for the current and voltage telemetry to travel from the DCIU back to the IAU. The additional round-trip delay is expected to be 3 clock cycles, or $\sim$300$\,$ms. To test this hypothesis a software simulation of the thrust controller and thruster response was performed using the flight thrust command data for CMNT\#5 from the period in Figure \ref{fig:CMNT5_alias}.  The green trace in Figure \ref{fig:CMNTsim} shows the simulated thrust error, sampled at 10Hz, for the nominal case of zero delay between the $I_B$ and $V_B$ commands and the corresponding response. This would be the case for the original flight configuration in which the thrust control algorithm was implemented on the DCIU. The light red trace shows the simulated thrust error, sampled at 10Hz, for the case where a 300$\,$ms delay is introduced between the command and response of the beam current and voltage.  This delay causes noise enhancement at  $\sim$700$\,$mHz by the control system, where the reduced control loop phase margin means that when the system attempts to suppress motion, the delayed commanded thrust actually mildly increases it instead, by the time the thrust change is enacted. The blue and yellow traces show the with- and without-delay signals, respectively, downsampled to 1Hz without anti-aliasing filters, as was done for nominal DRS operations. In both cases, this elevates the noise to roughly 50$\,\textrm{nN}/\sqrt{\textrm{Hz}}$ and, for the delay case, produces a sharp peak near 300$\,$mHz. 

An expanded  8-thruster version of this simulation, including modeling of the commanding loop delays and flight software, a physically motivated model of the bubble-noise on thruster 1, and the decimation scheme for creating the 1 Hz data product, provided an estimate of the expected platform noise. The data produced by this model matched the available mission flight data, both the 10 Hz data and the heavily aliased 1 Hz data. The 10 Hz data provided over a sufficient duration gave an estimate of the noise in the required frequency band, without the aliasing effects. This system noise along a particular direction was computed using knowledge of the CMNT locations and orientations. Using this simulation of the noise floor from the 10 Hz data, of $\sim$40$\,\textrm{nN}/\sqrt{\textrm{Hz}}$ for thrusters $2-8$ and 74$\,\textrm{nN}/\sqrt{\textrm{Hz}}$ for thruster 1, the estimated noise floor along spacecraft X,Y, and Z are $\sim$70$\,\textrm{nN}/\sqrt{\textrm{Hz}}$, $\sim$87$\,\textrm{nN}/\sqrt{\textrm{Hz}}$, and $\sim$56$\,\textrm{nN}/\sqrt{\textrm{Hz}}$.
   
\subsubsection{CMNT Noise Estimated from Spacecraft Response}
\label{sec:CMNTnoiseSC}
The procedure for estimating the thrust noise from the spacecraft response is very similar to that used for calibrating the thrusters, described in section \ref{sec:thustCal}, with the exceptions that no modulations are applied to the thrusters and that here we also analyze the Y- and Z- degrees of freedom.  This  analysis can be applied to any segment of data where injections are not present. Here we present results from four segments that are listed in Table \ref{tab:segments}.  These segments were chosen to span various configurations of control systems and thrusters so as to better distinguish the contribution of the propulsion system to the overall spacecraft noise. Segment I represents the default configuration for LTP, with the ESA-provided DFACS in control of the spacecraft using the cold-gas thruster system. Segment II represents the design configuration for DRS, with the DCS in controlling the spacecraft using all 8 CMNTs (and the ESA-provided cold gas thruster system on standby). Segment III is from a brief `joint operations' campaign in which the DFACS controlled the spacecraft with the CMNTs and the cold gas system was on standby. Finally, Segment IV represents the DRS configuration after the CMNT\#4 anomaly, with the DCS controlling the spacecraft using 7 CMNTs, and with the cold-gas system partially enabled as an out-of-loop static `crutch'. 

\begin{table}
\caption{Experiments used to assess thruster noise performance from the spacecraft response. For each experiment a controller, either the ESA-provided DFACS or the NASA-provided DCS, controlled the spacecraft using a micropropulsion system, either the ESA-provided cold-gas (CGAS) or the NASA-provided CMNT. For segment IV, the DCS controlled the spacecraft with 7 CMNTs in-loop and the CGAS used as an out-of-loop `crutch'.\label{tab:segments}}
\begin{center}
\begin{tabular}{|c|c|c|c|c|}
\hline
Segment & Start Date & Duration & Controller & Propulsion \\
\hline
I & 2016-9-28 & 238 ks& DFACS & CGAS \\ \hline
II & 2016-10-04 & 111 ks& DCS & CMNT \\ \hline
III & 2016-10-06 & 124 ks& DFACS & CMNT \\ \hline
IV & 2017-04-21 & 236 ks& DCS & CMNT  \\ 
& & & & w/ CGAS \\
\hline
\end{tabular}
\end{center}
\end{table}
 
\begin{figure}
\begin{center}
\includegraphics[width=\columnwidth]{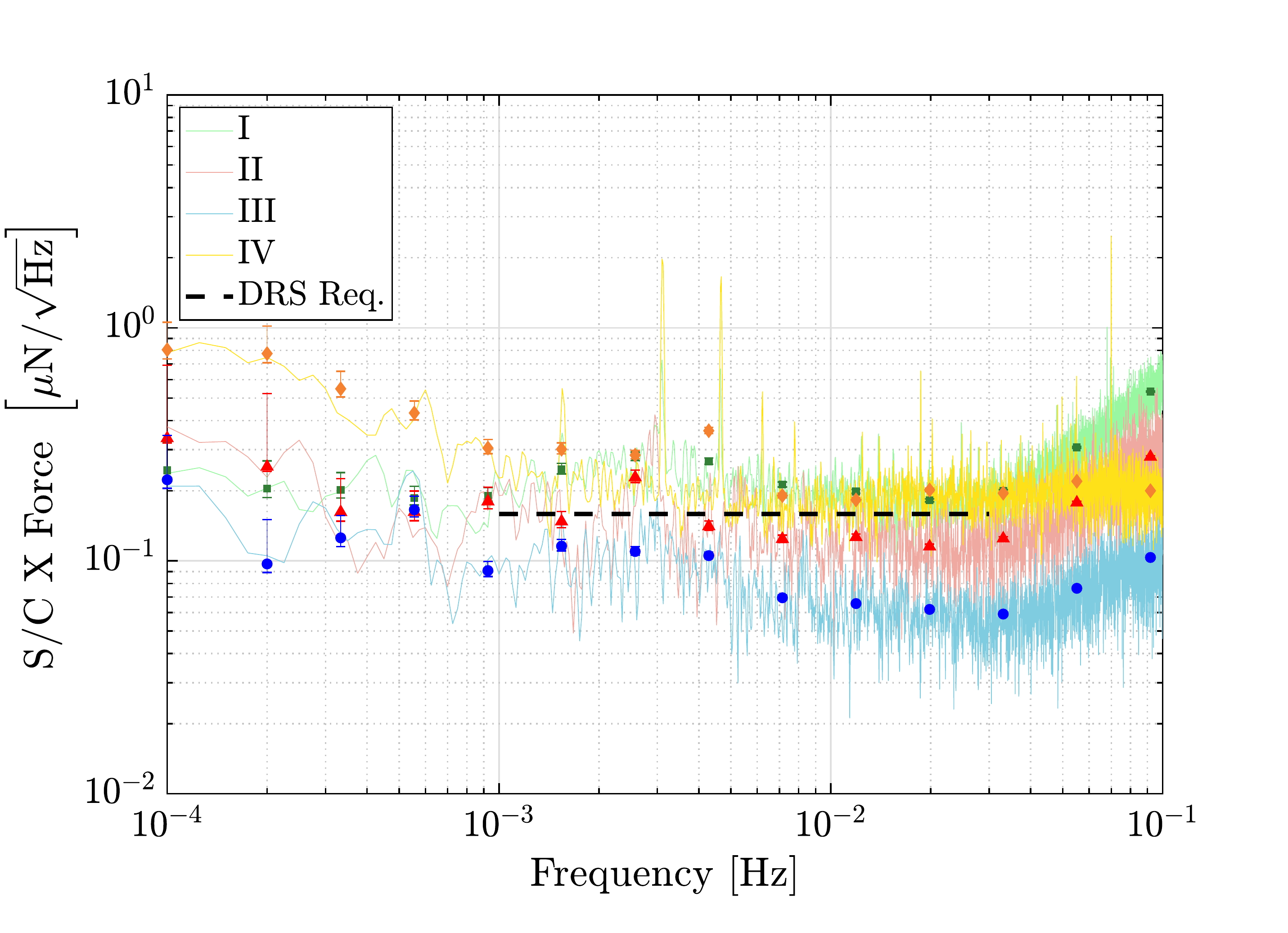}
\caption{X-component of force on the spacecraft estimated using measured test mass dynamics and spacecraft mass properties. The four traces correspond to the segments in \ref{tab:segments} and probe different thruster configurations. See text for discussion. }
\label{fig:Fext_x}
\end{center}
\end{figure}

\begin{figure}
\begin{center}
\includegraphics[width=\columnwidth]{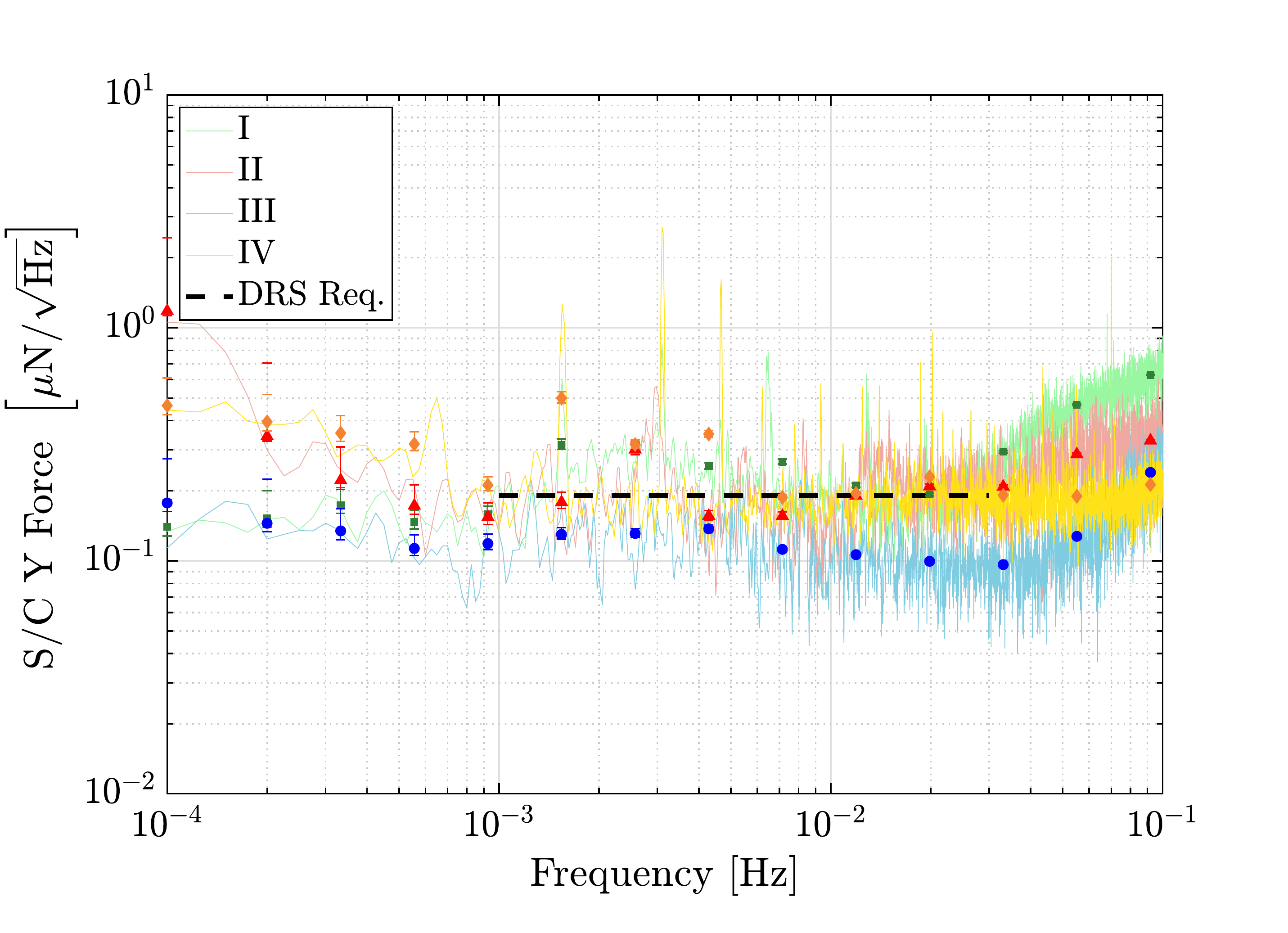}
\caption{Y-component of force on the spacecraft estimated using measured test mass dynamics and spacecraft mass properties. The four traces correspond to the segments in \ref{tab:segments} and probe different thruster configurations. See text for discussion. }
\label{fig:Fext_y}
\end{center}
\end{figure}

\begin{figure}
\begin{center}
\includegraphics[width=\columnwidth]{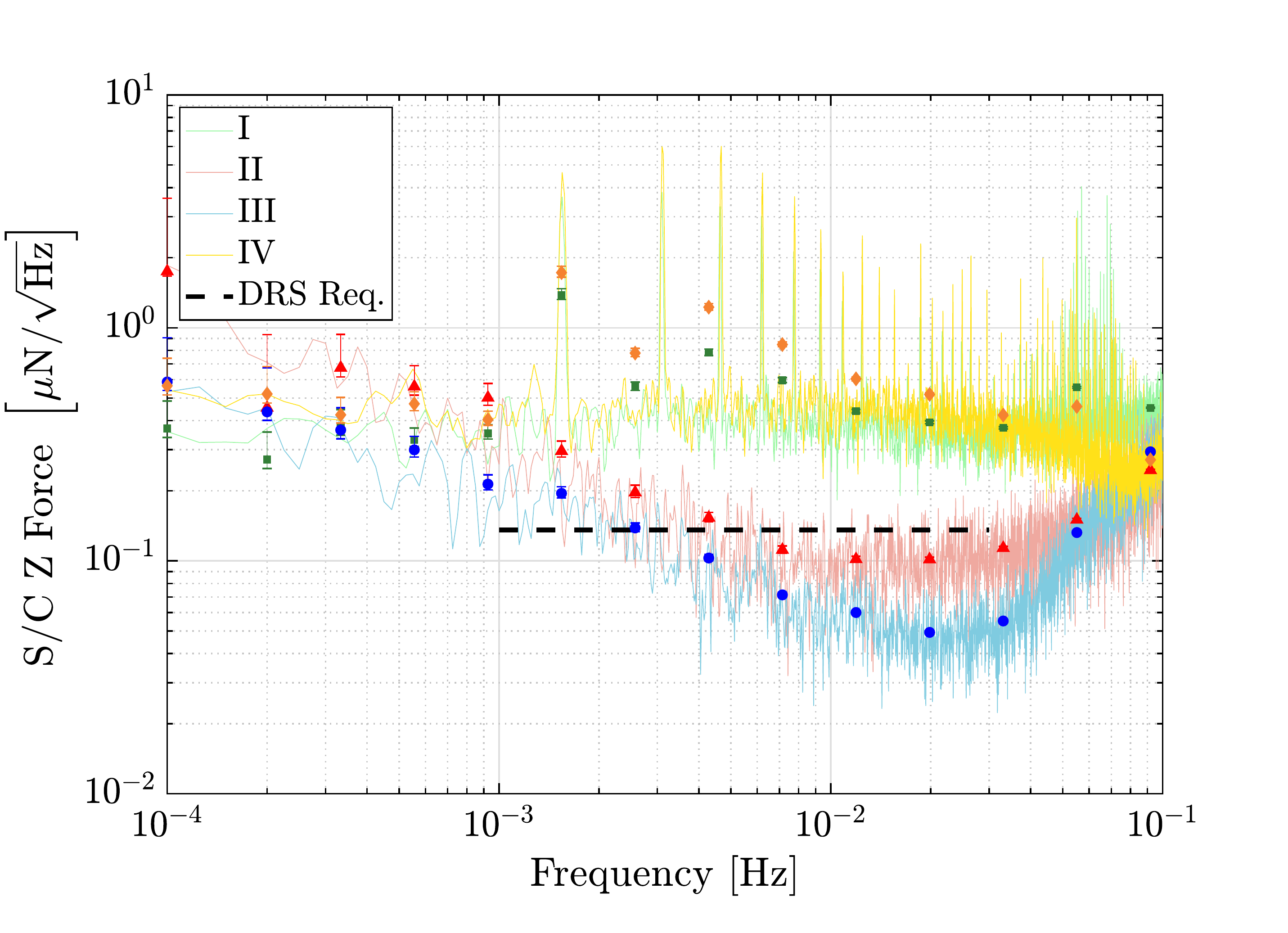}
\caption{Z-component of force on the spacecraft estimated using measured test mass dynamics and spacecraft mass properties. The four traces correspond to the segments in \ref{tab:segments} and probe different thruster configurations. See text for discussion. }
\label{fig:Fext_z}
\end{center}
\end{figure}

 Figures \ref{fig:Fext_x}, \ref{fig:Fext_y}, and \ref{fig:Fext_z} show the estimated spacecraft force noise in the X-, Y-, and Z-directions for each of the four segments. The solid lines are amplitude spectral densities estimated using Welch's method of overlapped-averaged periodograms with 10,4,5, and 10 averages for segments I,II,II, and IV, respectively. For segment III, an impulse suspected to be from a micrometeoroid hit at 2016-10-07 9:51:12 UTC was excised from the data.  The solid points represent logrithmically-binned estimates with 1-sigma error bars. The solid dashed lines show a `requirement' based on an uncorrelated thrust noise of 0.1$\,\mu\textrm{N}/\sqrt{\textrm{Hz}}$ in each of the CMNTs projected into the spacecraft body frame using the thruster orientations. This corresponds to  160$\,\textrm{nN}/\sqrt{\textrm{Hz}}$,  190$\,\textrm{nN}/\sqrt{\textrm{Hz}}$, and 140$\,\textrm{nN}/\sqrt{\textrm{Hz}}$ for X-, Y-, and Z-axes respectively.
 
 In general, the noise for the two CMNT-only configurations (II and III) are somewhat lower than the two including cold gas (I and IV). In addition to the overall higher noise level, the two cold-gas segments exhibit a set of narrow-line features at $\sim1.5\,$mHz and harmonics thereof.  Both of these effects are most pronounced in the Z-axis, possibly explained by a common-mode noise source in the cold gas system\footnote{ Since all 6 Cold Gas thrusters thrust in the +Z (Sunward) direction with the same vector component, a common-mode noise will add coherently whereas correlated noise in X and Y would largely cancel when all 6 thrusters are active. Note that for the case of the `crutch' mode using only 4 of the 6 thrusters, the cancellation in X and Y no longer occurs.} Somewhat surprisingly, the CMNT noise under DFACS control (III) appears to be slightly lower than that for DCS (II). Upon inspection of the telemetry it was found that CMNT\#1 was railed at the minimum thrust of 5$\,\mu$N due to a unoptimized thrust bias vector for this ad-hoc experiment. This inadvertently reduced the rate of `blipping' in CMNT\#1, leading to a reduction in the overall noise. 
 
 \subsubsection{CMNT Noise Summary}
 
In both CMNT-only cases, the measured noise floor along the x-direction seems to be in the $100\sim 300\,\textrm{nN}/\sqrt{\textrm{Hz}}$ range, which is significantly higher than the noise predicted by the thrust telemetry in the absence of aliasing. Possible explanations for this include additional noise in the thrust system beyond what is inferred from the current and voltage noise, additional noise on the spacecraft platform, or noise on the test mass which is used as a reference. 

 \begin{table}
\caption{Comparison of measured spacecraft Force noise  and combined estimate from measured and estimated effects in the in $1\sim30\,$mHz band. Details of noise estimates can be found in Appendix C. \label{tab:SCnoiseComp} }
\begin{center}
\begin{tabular}{|c|c|}
\hline
Effect & Estimate nN$/\sqrt{\textrm{Hz}}$ \\
\hline
Measured Noise & 120  \\
\hline
Total Estimate & 70  \\
\hline
Unmodeled noise & 97  \\
\hline
\multicolumn{2}{|c|}{\emph{CMNT noises}} \\
\hline
IV noise (CMNT\#1) & 41\\
IV noise (CMNT\#2-7) & 56 \\
shot noise &  0.16 \\
flutter  noise & 0.03 \\
\hline
\multicolumn{2}{|c|}{\emph{S/C noises}} \\
\hline
SRP & 1.7 \\
Radiometer Noise & 0.7 \\
Ext. B-fields & 0.01 \\
Micrometeoroids & 0.5 \\
\hline
\multicolumn{2}{|c|}{\emph{TM noises}} \\
\hline
force noise & $<$ 1 \\
\hline
\end{tabular}
\end{center}
\vspace{0cm}
\end{table}%

Table \ref{tab:SCnoiseComp} summarizes measurements and estimates of some of these effects for the x-axis of the spacecraft in the $1\sim30\,$mHz band. Details on how some of these effects were estimated can be found in Appendix C. Measured Noise is an approximate white noise level equivalent to the red trace (segment II) in Figure \ref{fig:Fext_x}. Total Estimate is a uncorrelated sum of the remaining entries in the table, which represents the total amount of noise accounted for in our model. This is dominated by an estimate of the modeled thrust noise  without the presence of aliasing that is estimated from the simulations of current and voltage and resulting noise floor presented in section \ref{sec:IVnoise} and Figure \ref{fig:CMNTsim}. The table lists the contribution from CMNT\#1, which has an elevated noise floor due to the blipping, as well as the sum of the rest of the thrusters. Unmodeled Noise is the size of the noise contribution, presumably uncorrelated with the modeled noise, that would be needed to be added to the model to match our measurements. This represents more than half of the measured noise, but at this time we are unable to account for the source of this effect. This suggests that designers of future low-disturbance platforms should take care when considering applications requiring force noise below $\sim100\,\textrm{nN}/\sqrt{\textrm{Hz}}$.
\\

\section{DRS Performance}
\subsection{Operation of the Dynamic Control System}
\begin{figure*}\label{transitions}
\centering
\begin{tabular}{ccc}

\subf{\includegraphics[width=0.32\textwidth]{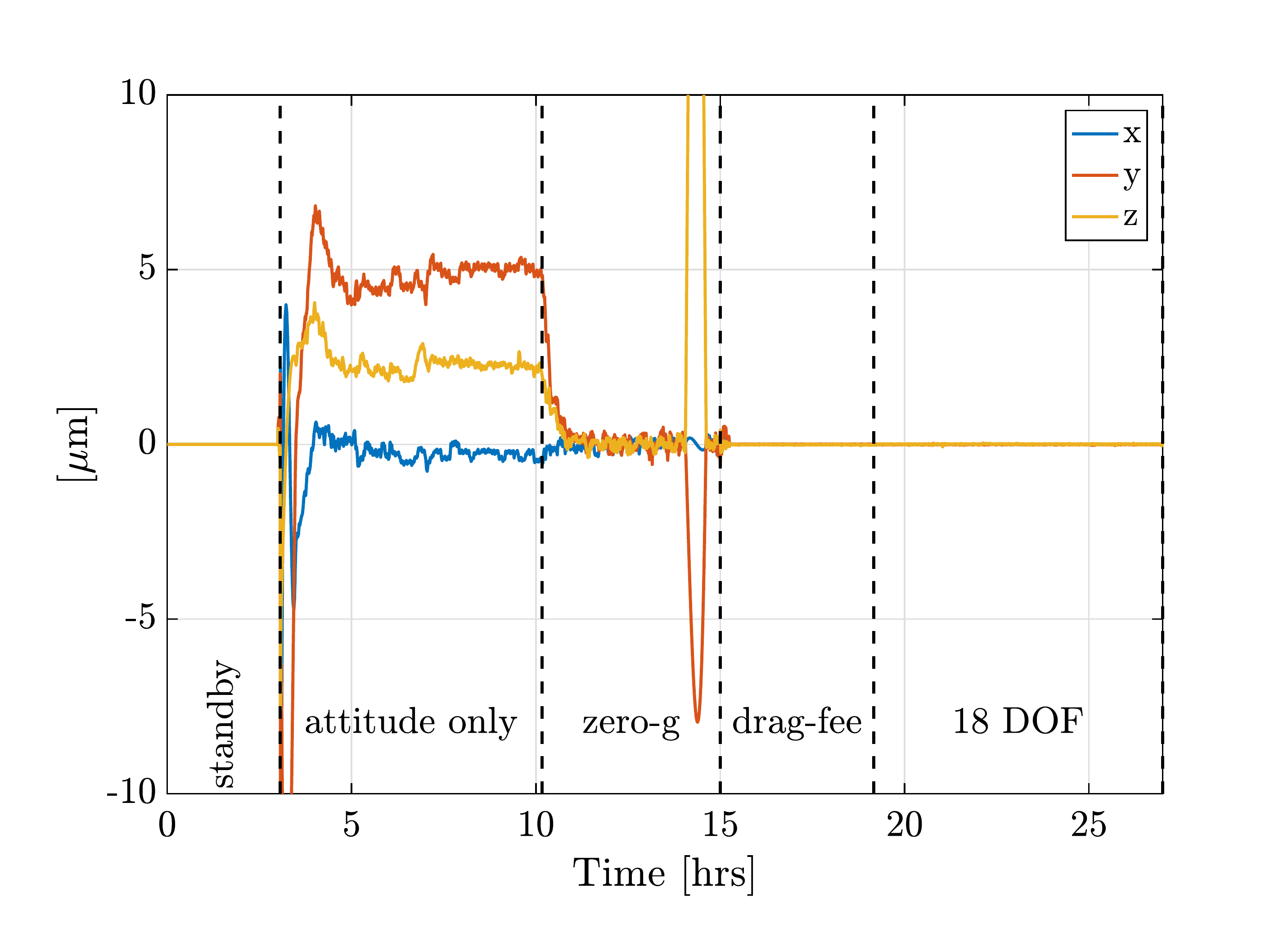}}
     {RTM positions}
&
\subf{\includegraphics[width=0.32\textwidth]{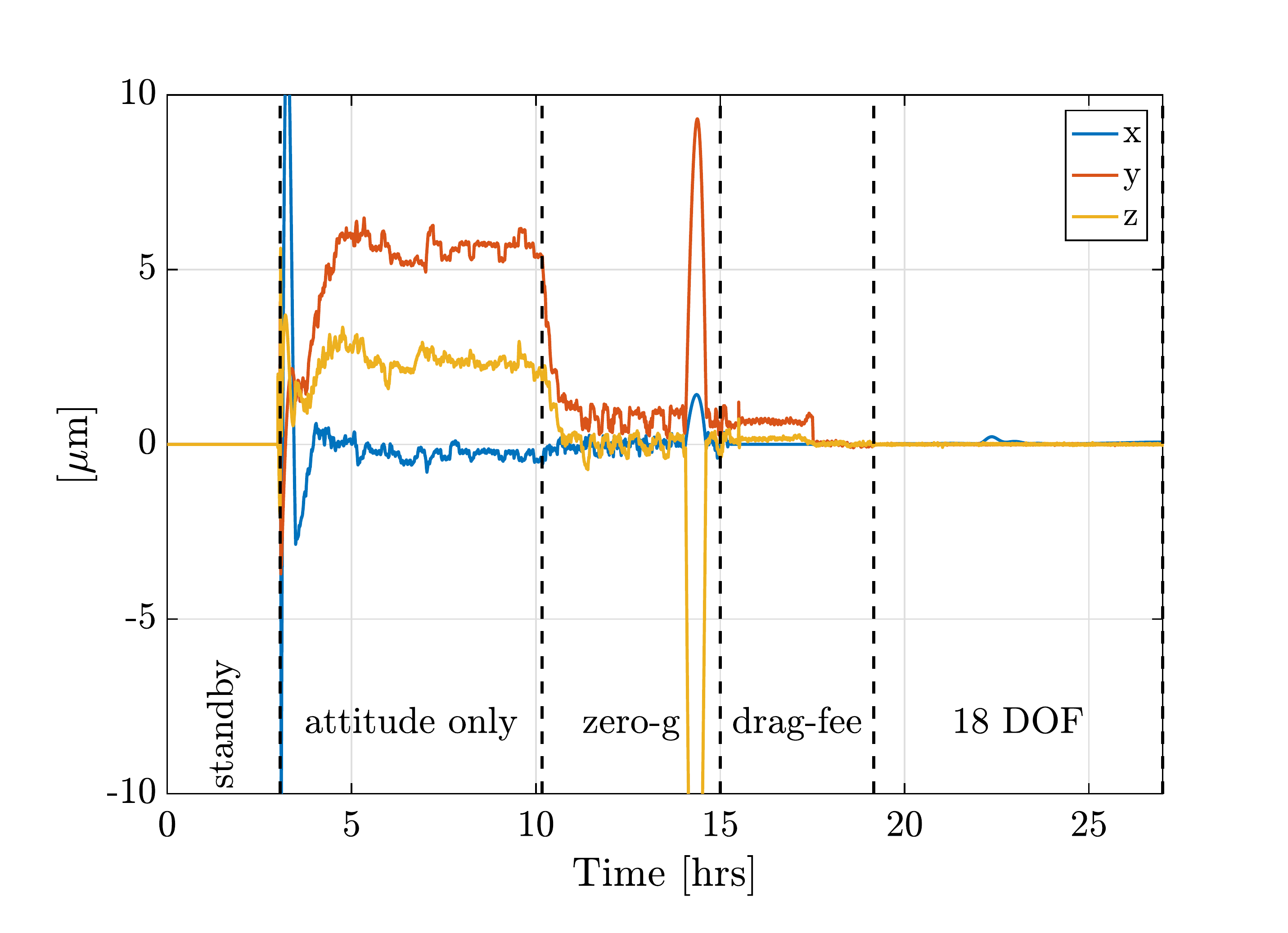}}
     {NTM positions}
 &
\\
\subf{\includegraphics[width=0.32\textwidth]{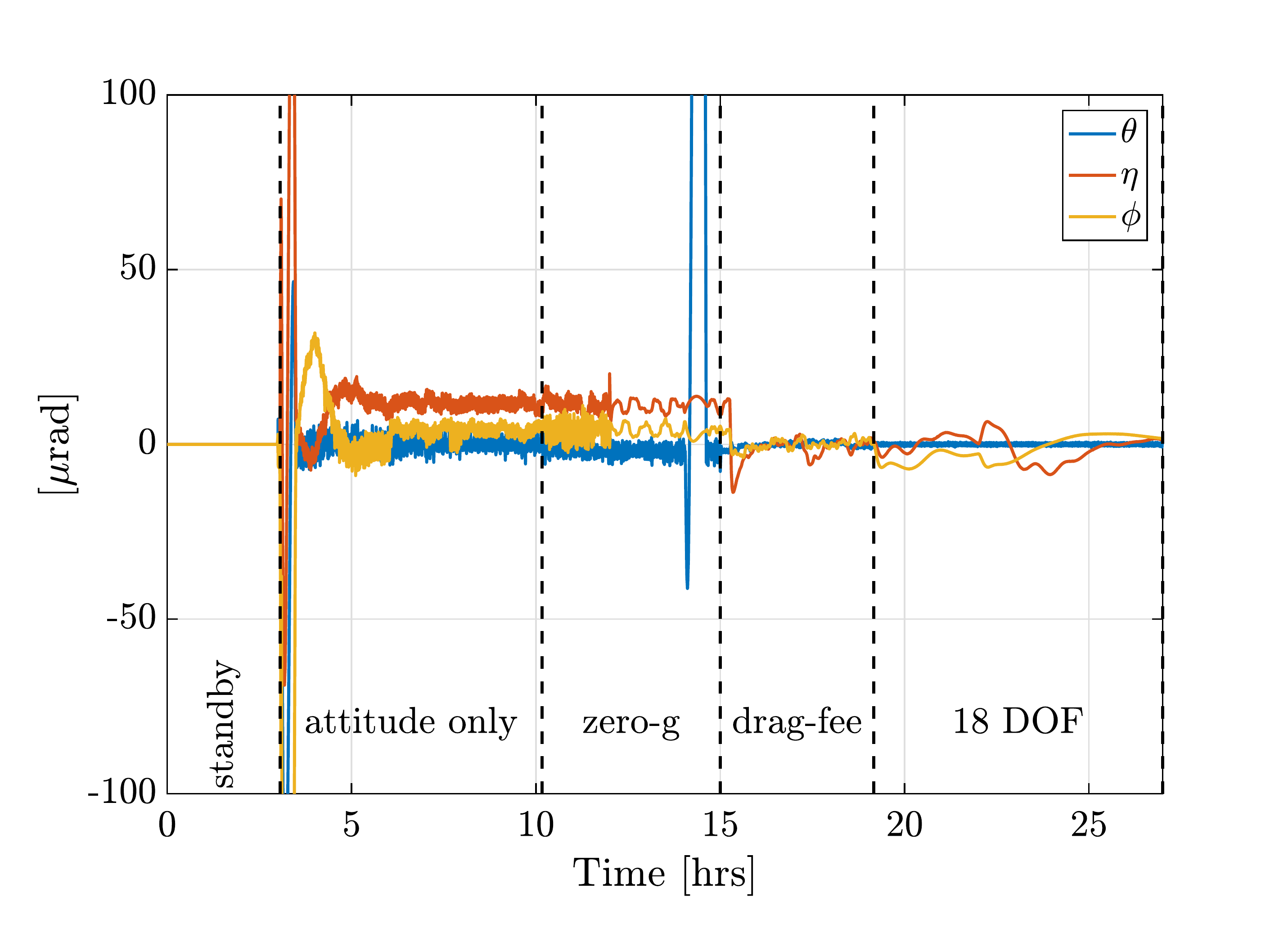}}
     {RTM angles}
&
\subf{\includegraphics[width=0.32\textwidth]{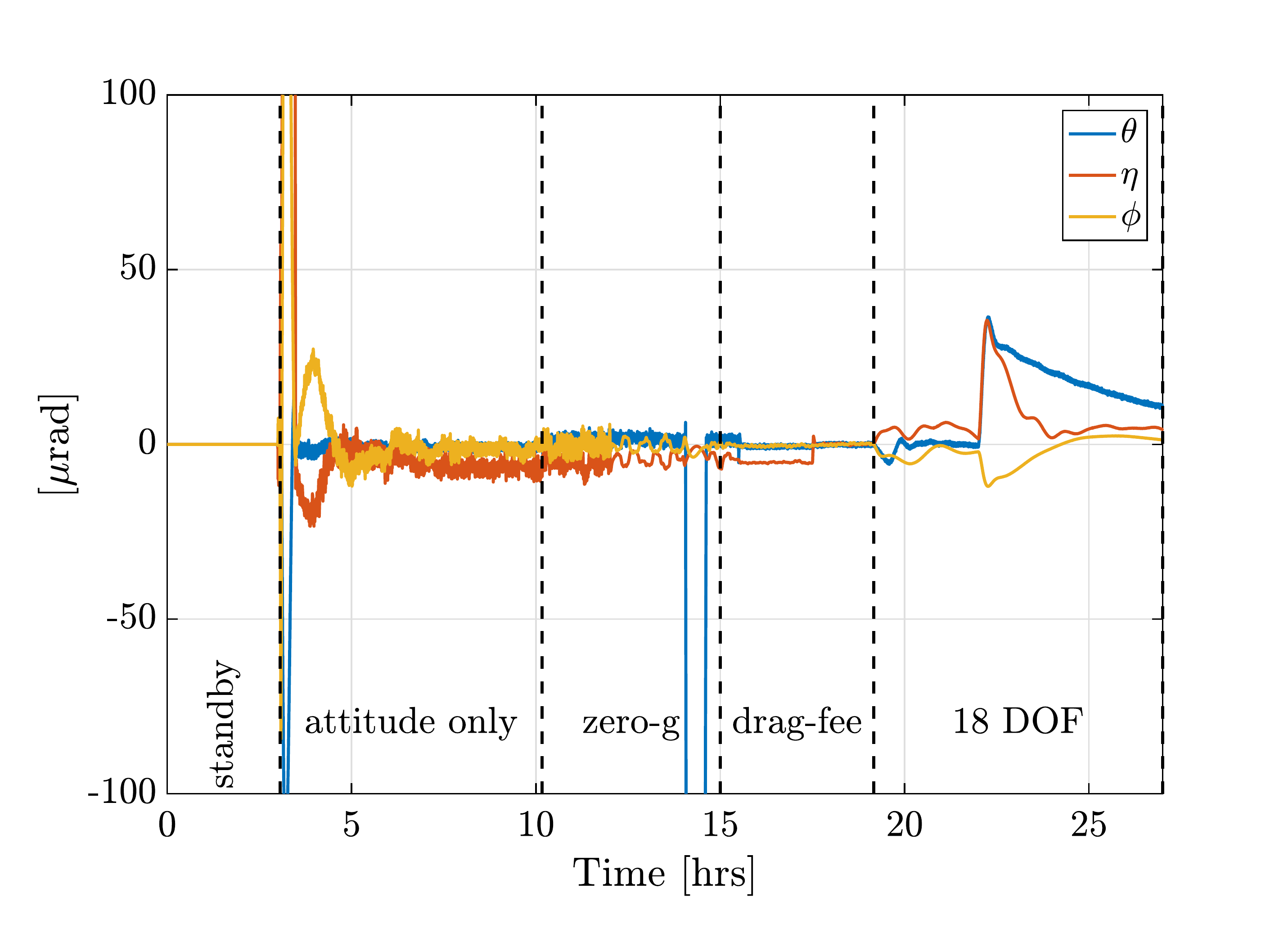}}
     {NTM angles}
 &
 \subf{\includegraphics[width=0.32\textwidth]{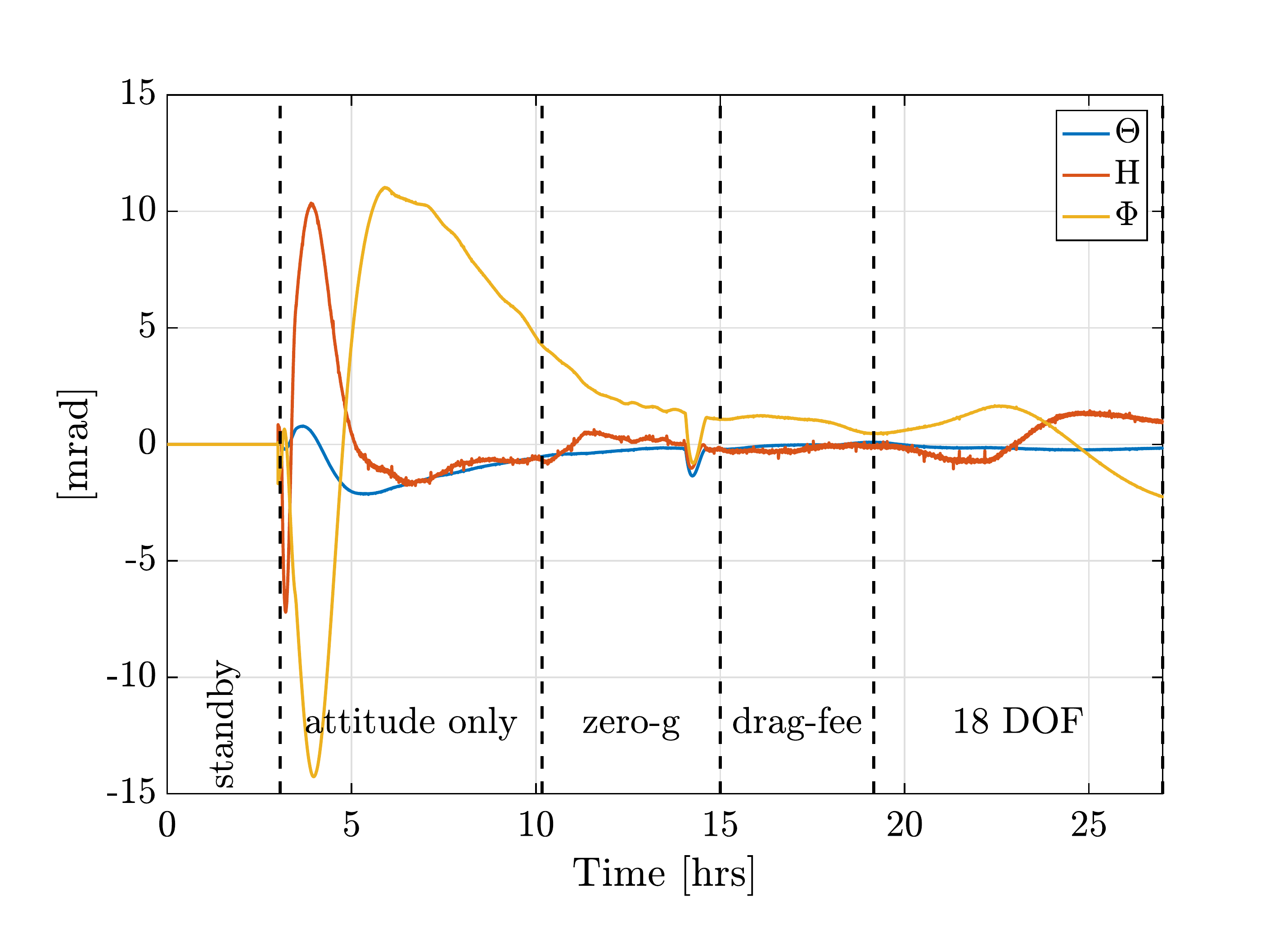}}
     {SC angles}
\\
\subf{\includegraphics[width=0.32\textwidth]{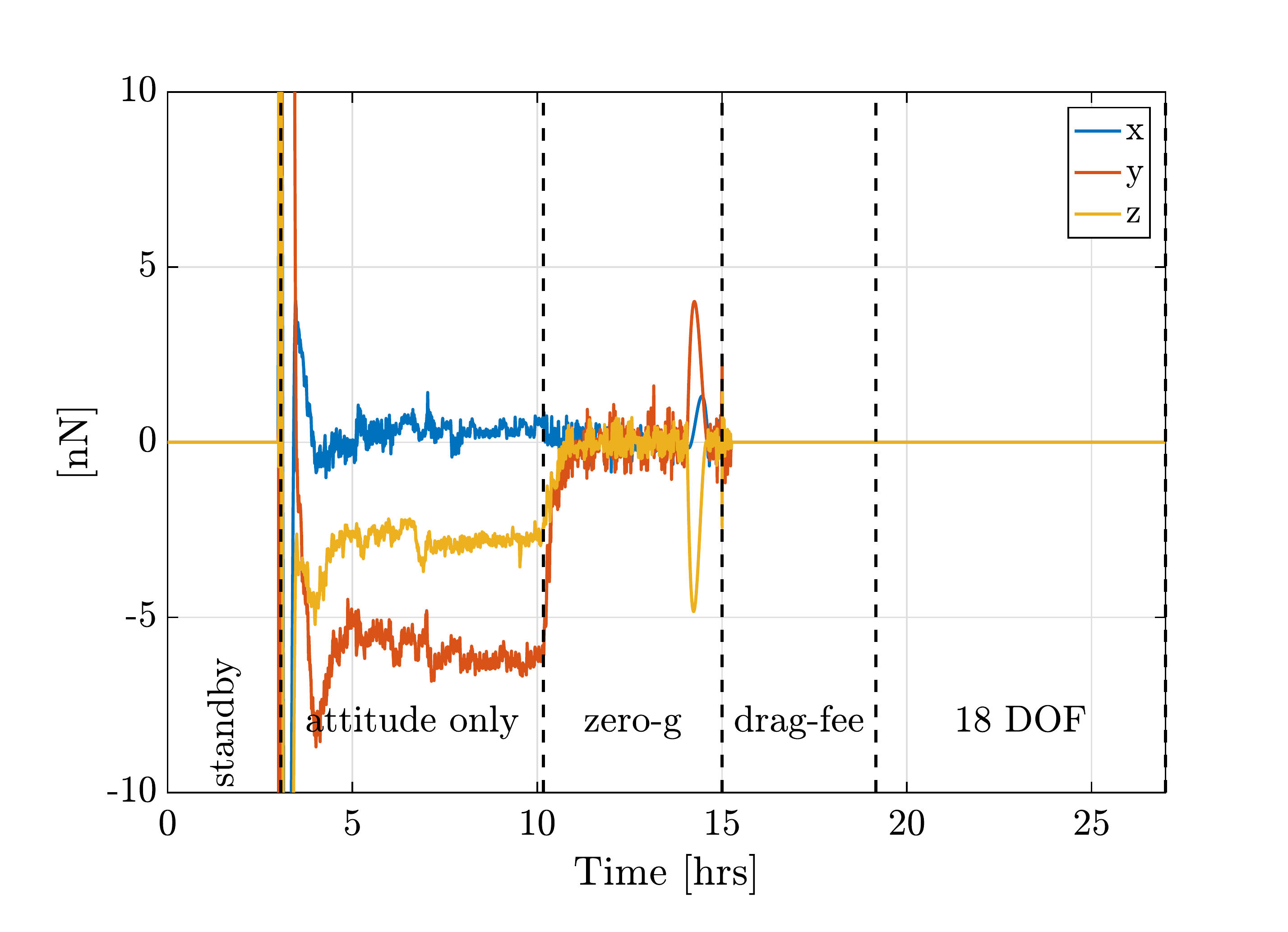}}
     {RTM forces}
&
\subf{\includegraphics[width=0.32\textwidth]{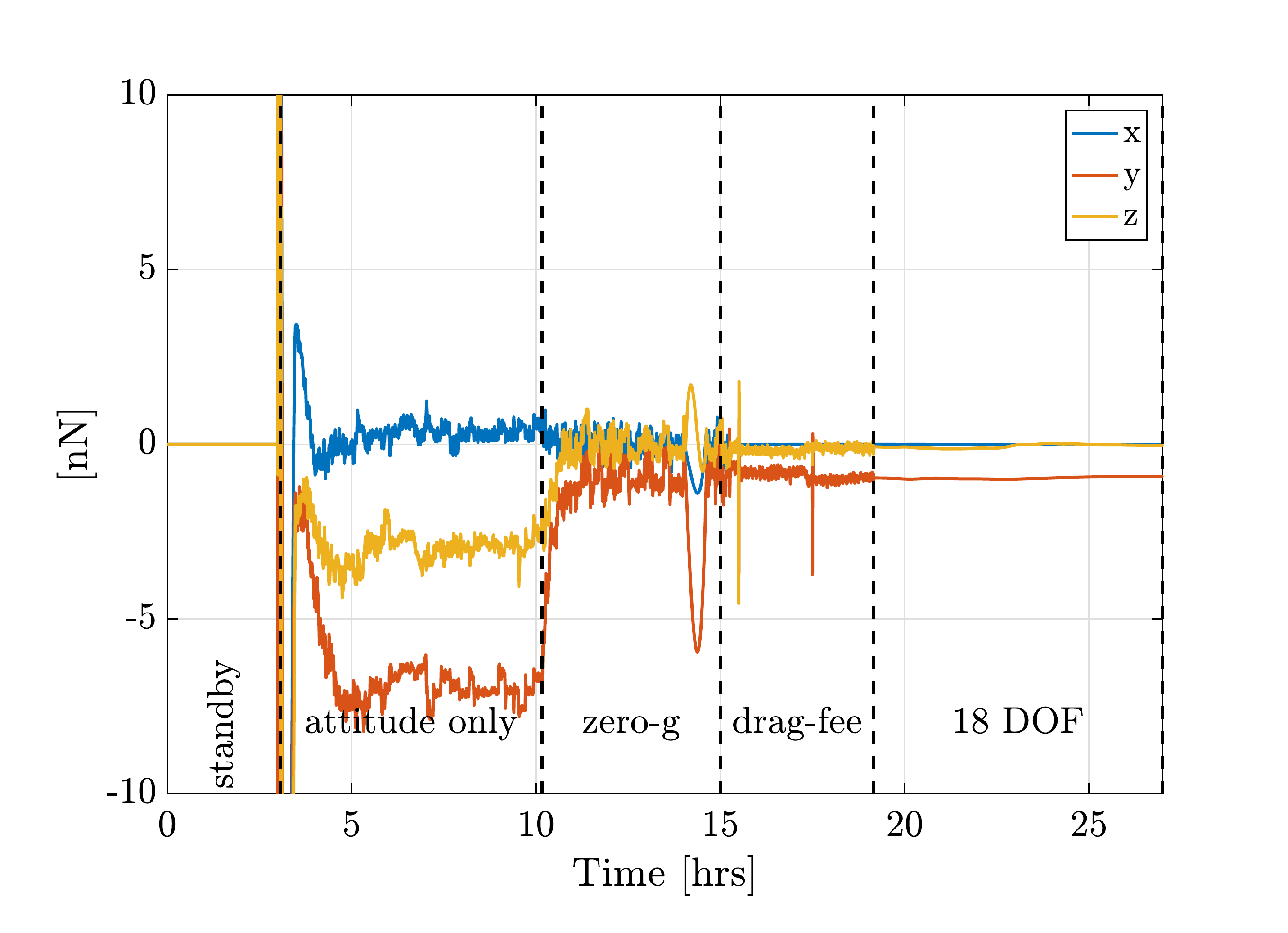}}
     {NTM forces}
 &
 \subf{\includegraphics[width=0.32\textwidth]{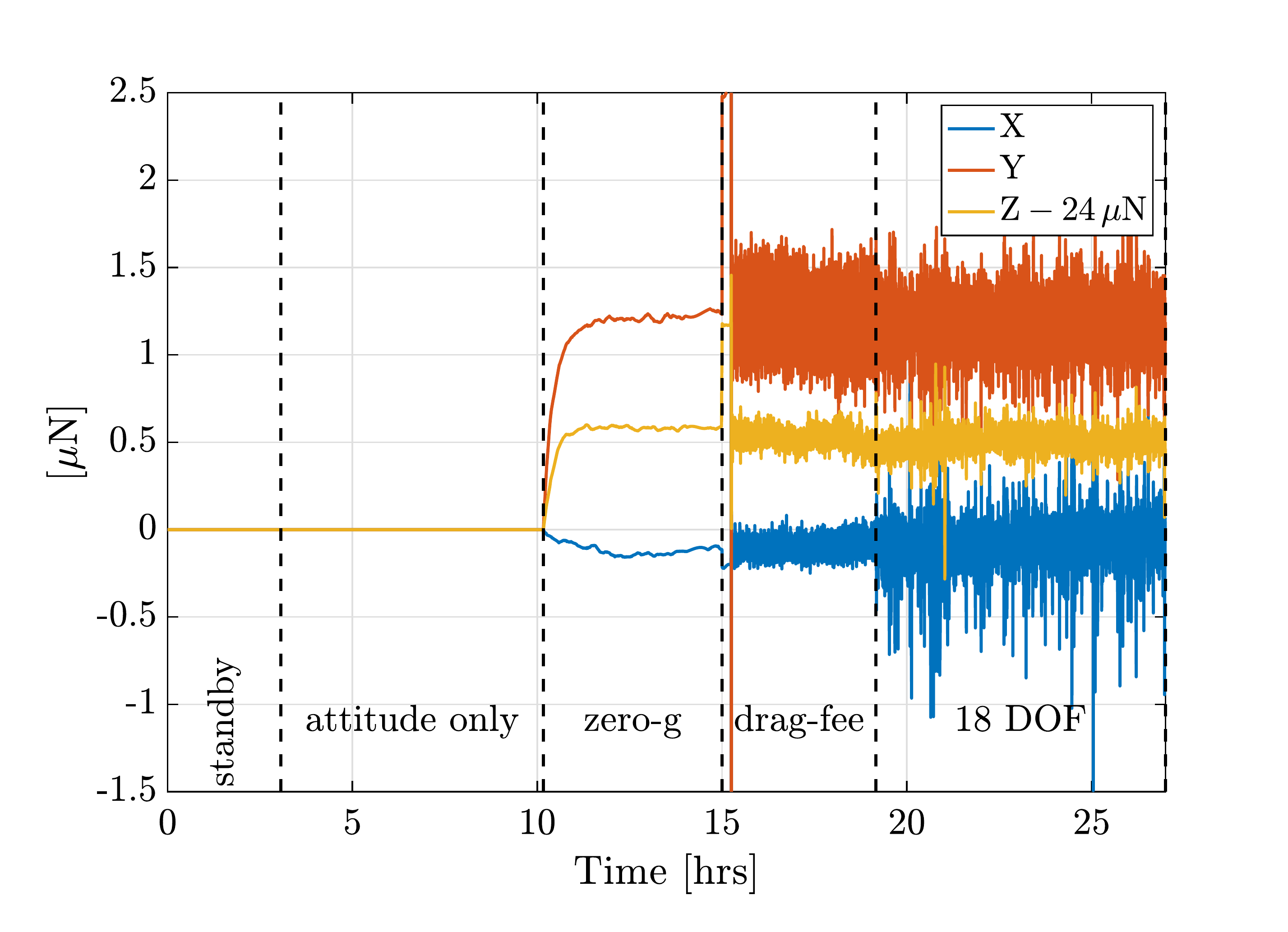}}
     {SC forces}
\\
\subf{\includegraphics[width=0.32\textwidth]{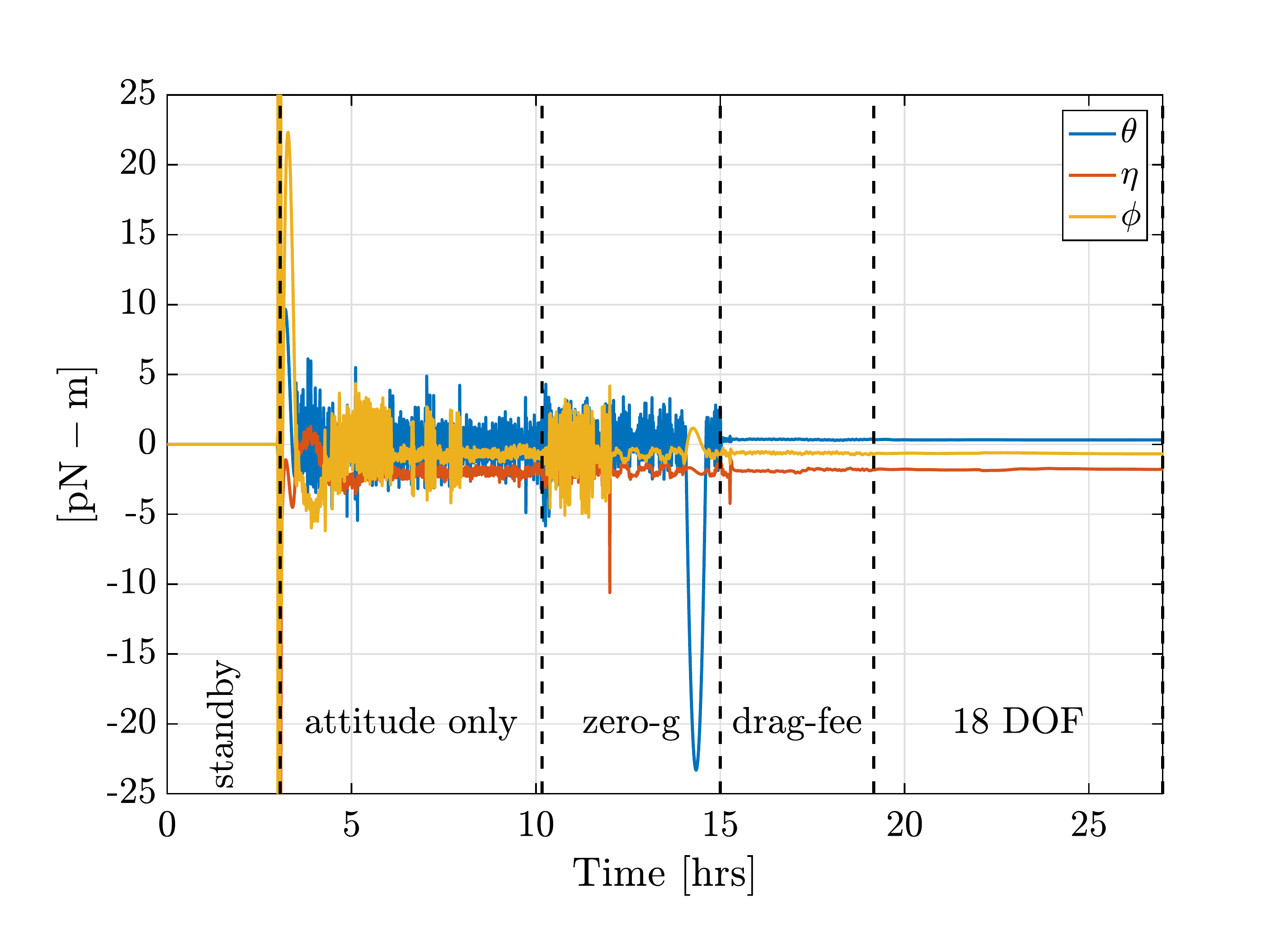}}
     {RTM torques}
&
\subf{\includegraphics[width=0.32\textwidth]{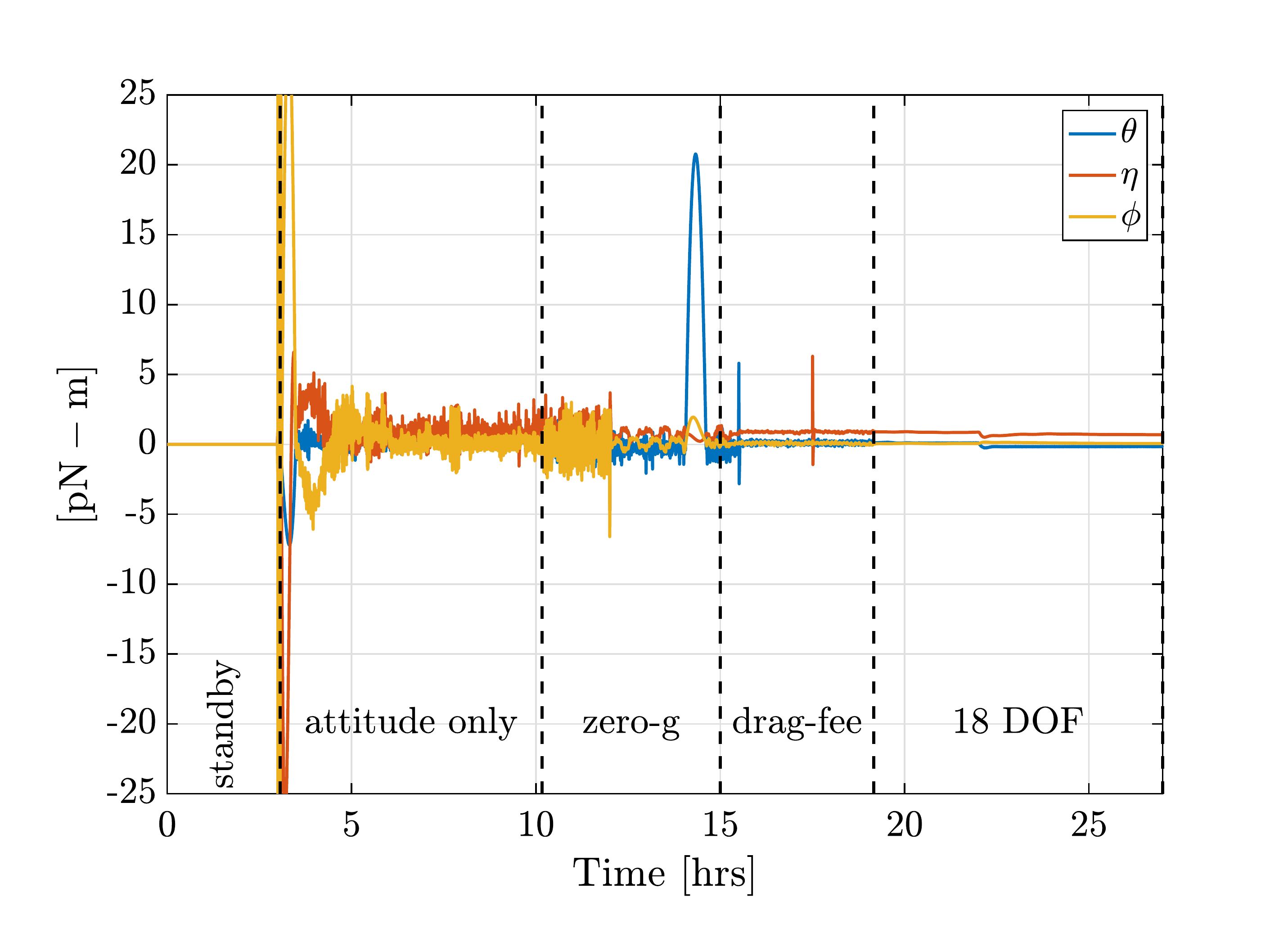}}
     {NTM torques}
 &
 \subf{\includegraphics[width=0.32\textwidth]{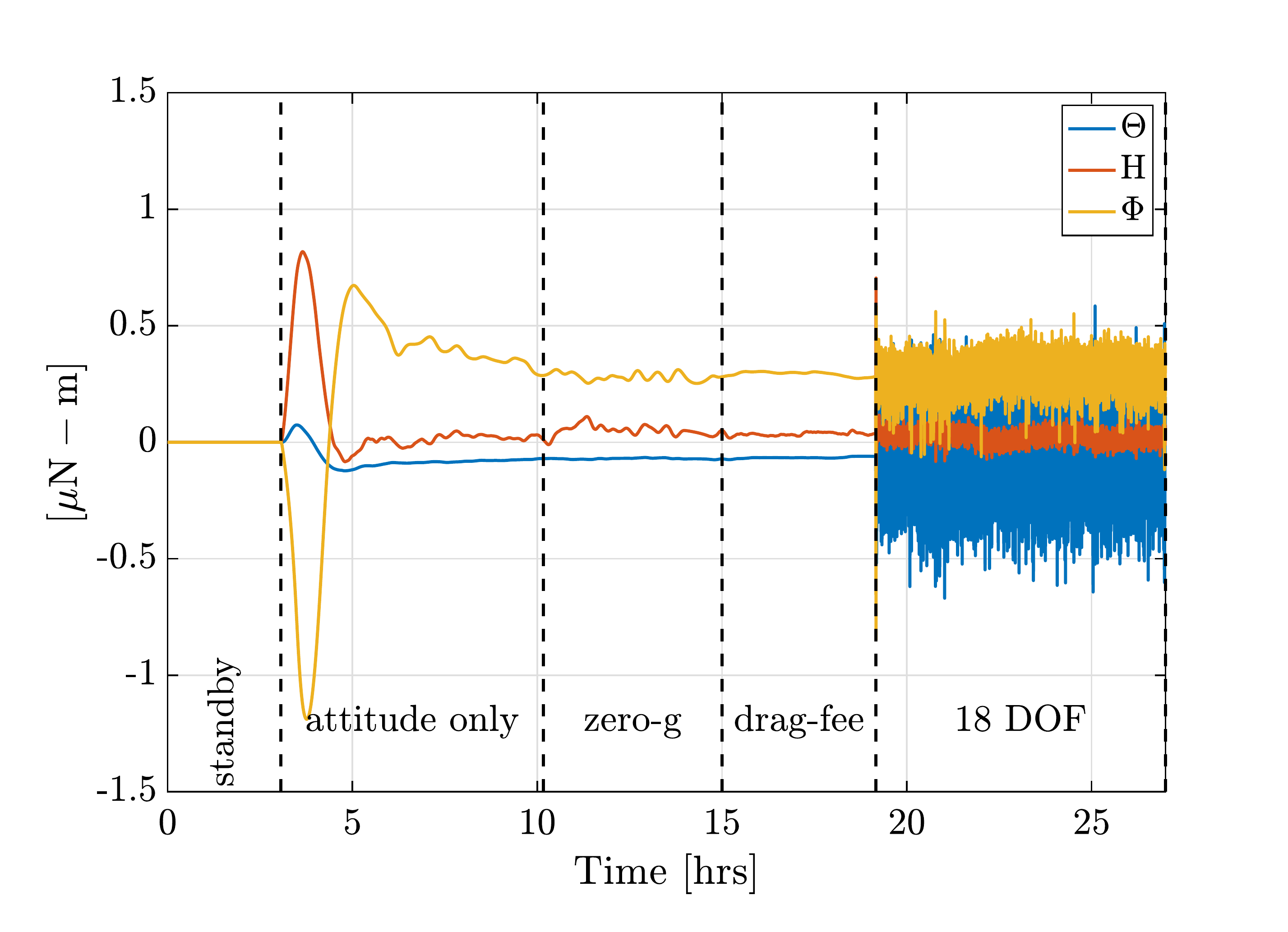}}
     {SC torques}
\\
\end{tabular}
\caption{DCS behavior during a typical mode transition sequence from handover to the 18 degree-of-freedom science mode (18DOF). Plots show measured positions and angles of both the reference (RTM) and non-reference (NTM) test masses; angles of the spacecraft; and forces and torques applied to the RTM, NTM, and spacecraft . Note that for spacecraft forces, the thruster bias levels are set to provide a net force in the $+Z$ direction of 24$\,\mu$N. Time origin is 2016-10-02 14:00UTC.}
\label{fig:modeTransitions} 
\vspace{2cm}
\end{figure*}

DRS operations were initialized with a handover sequence wherein control of the spacecraft and test masses was passed from the European DFACS control system to the DCS.  After initial capture, the DCS executed sequences to transition through the various control modes described in section \ref{sec:DCS} in order to bring the instrument to the desired state for conducting experiments. During DRS operations, this procedure occurred roughly once per week after station-keeping maneuvers were performed under DFACS control. Figure~\ref{fig:modeTransitions} shows the measured positions and attitudes as well as the commanded forces and torques of the test mass and spacecraft for a typical transition sequence from handover to the 18DOF science mode. Note that spacecraft forces do not include the bias levels of the thrusters, which were adjusted to provide a net DC force of 24$\,\mu$N in the $+Z$ direction to compensate for solar radiation pressure. After an initial transient caused by the handover sequence, the Attitude Only controller works to stabilize the spacecraft angular error. Both test masses are commanded to follow the spacecraft by applying appropriate forces and torques.  For the Zero-g mode, forces are applied to the spacecraft to minimize the z-axis forces on the test masses, thereby providing active compensation of the solar radiation pressure. In the drag-free mode, the spacecraft is commanded to follow the RTM along the linear DoFs, leading to an increase in the applied forces on the spacecraft and a decrease in both the position error and applied force on the RTM. The 18DOF mode, torques are applied to the spacecraft to further reduce the forces on the NTM in the transverse directions. In this particular sequence, the NTM experienced an impulsive disturbance approximately 2.5 hours after the transition into 18DOF mode that caused an excursion of the NTM angles and x position. The DCS compensated for this disturbance by applying appropriate torques and forces to the NTM. The DCS successfully executed dozens of mode transition sequences over the course of the baseline and extended mission, providing a robust platform with which to conduct experiments characterizing the CMNTs and other aspects of the spacecraft.

\subsection{Position Accuracy}
\begin{figure}
\begin{center}
\includegraphics[width=\columnwidth]{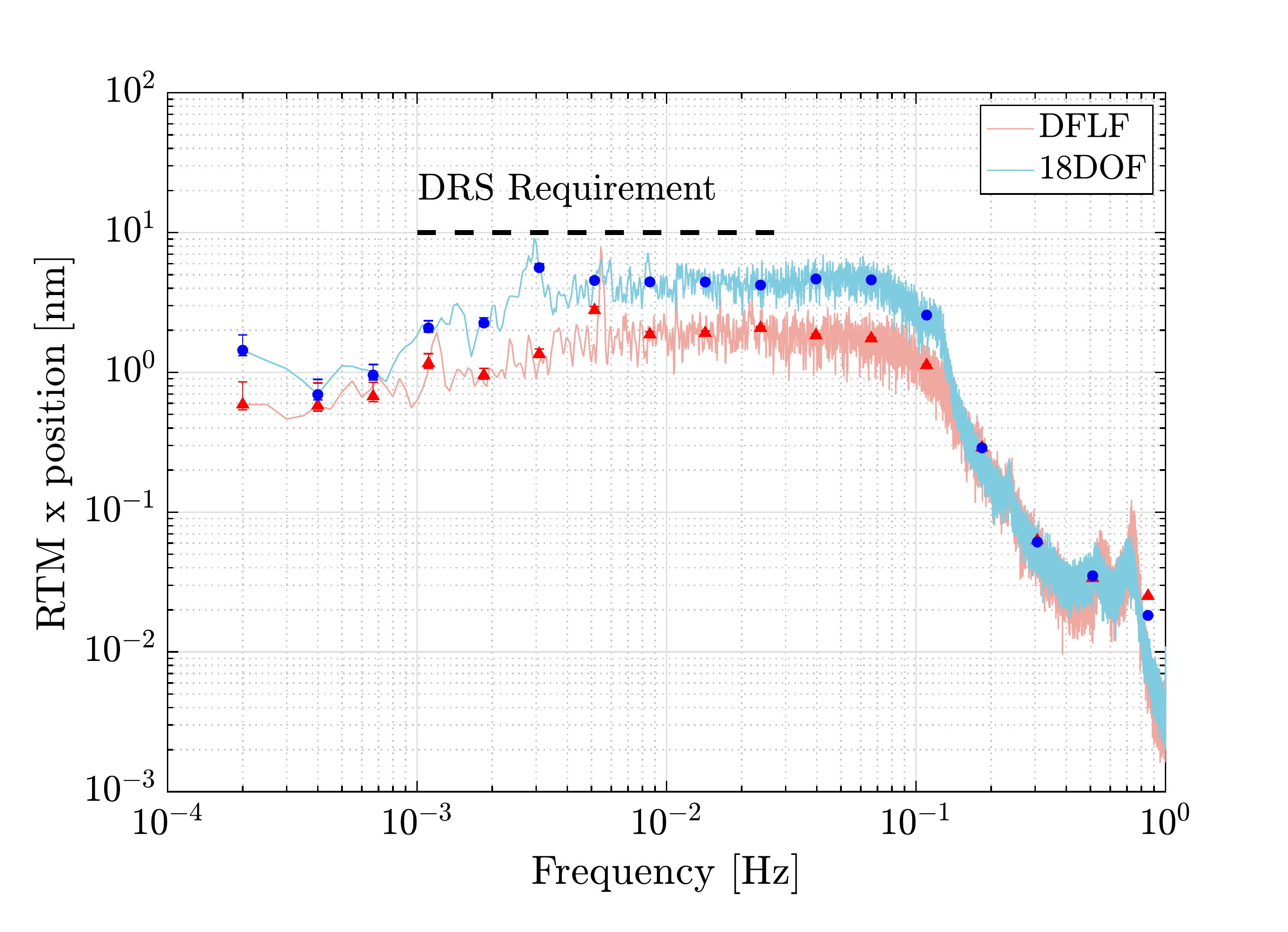}
\caption{Amplitude spectral density of the measured RTM-Sspacecraft position along the x-direction for a 20.7 hr run in the  drag-free low-force (DFLF, red) mode beginning on 2016-08-22 and a 31.1 hr run in the 18 degree-of-freedom (18DOF, blue) mode beginning on 2016-10-22. The solid trace shows a linearly-binned spectral density computed using Welch's method with a frequency resolution of 25$\,\mu$Hz while the solid points show a logarithmically-binned estimate with 1-sigma error bars. The black dashed line is the Level I position error requirement for ST7-DRS.}
 \label{fig:position_error_f} 
\label{default}
\end{center}
\end{figure}

The spacecraft position error, or the precision with which the spacecraft position is maintained relative to the RTM, is an important requirement for the DCS. The ST7-DRS Level I requirement was for a position error amplitude spectral density of $S_{SCx}^{1/2}\leq10\,\textrm{nm}/\sqrt{\textrm{Hz}}$ in the band $1\,\textrm{mHz}\leq f \leq 30\,\textrm{mHz}$. Figure~\ref{fig:position_error_f} shows the measured  $S_{SCx}^{1/2}$ for two different experiments: a 20.7 hr run in the drag-free low-force (DFLF) mode beginning on 2016-08-22 07:36 UTC and a 31.1 hr run in the 18 degree-of-freedom (18DOF) mode beginning on 2016-10-22 00:00 UTC.  Both control modes comfortably meet the requirement over the measurement band. At high frequencies, both traces show a spectrum that follows a roughly $f^{-2}$ power-law and has an amplitude that is consistent with a white force noise on the order of $\sim0.1\,\mu\textrm{N}/\sqrt{\textrm{Hz}}$ and a spacecraft mass of 422$\,$kg.  As discussed in section \ref{sec:CMNTnoise}, thruster noise on the order of $\sim0.1\,\mu\textrm{N}/\sqrt{\textrm{Hz}}$ is expected to dominate the spacecraft force noise budget. At $\sim100\,$mHz,  the drag-free controller begins compensating for this disturbance by applying force commands on the spacecraft , leading to a flattening of the position error spectrum. The DFLF controller has a slightly higher control bandwidth for the x-axis drag-free loop than the 18DOF controller, resulting in a slightly lower level of position noise within the control bandwidth.  At lower frequencies, additional gain in the drag-free loop further suppresses disturbances from the thrusters.

\begin{figure}
\begin{center}
\includegraphics[width=\columnwidth]{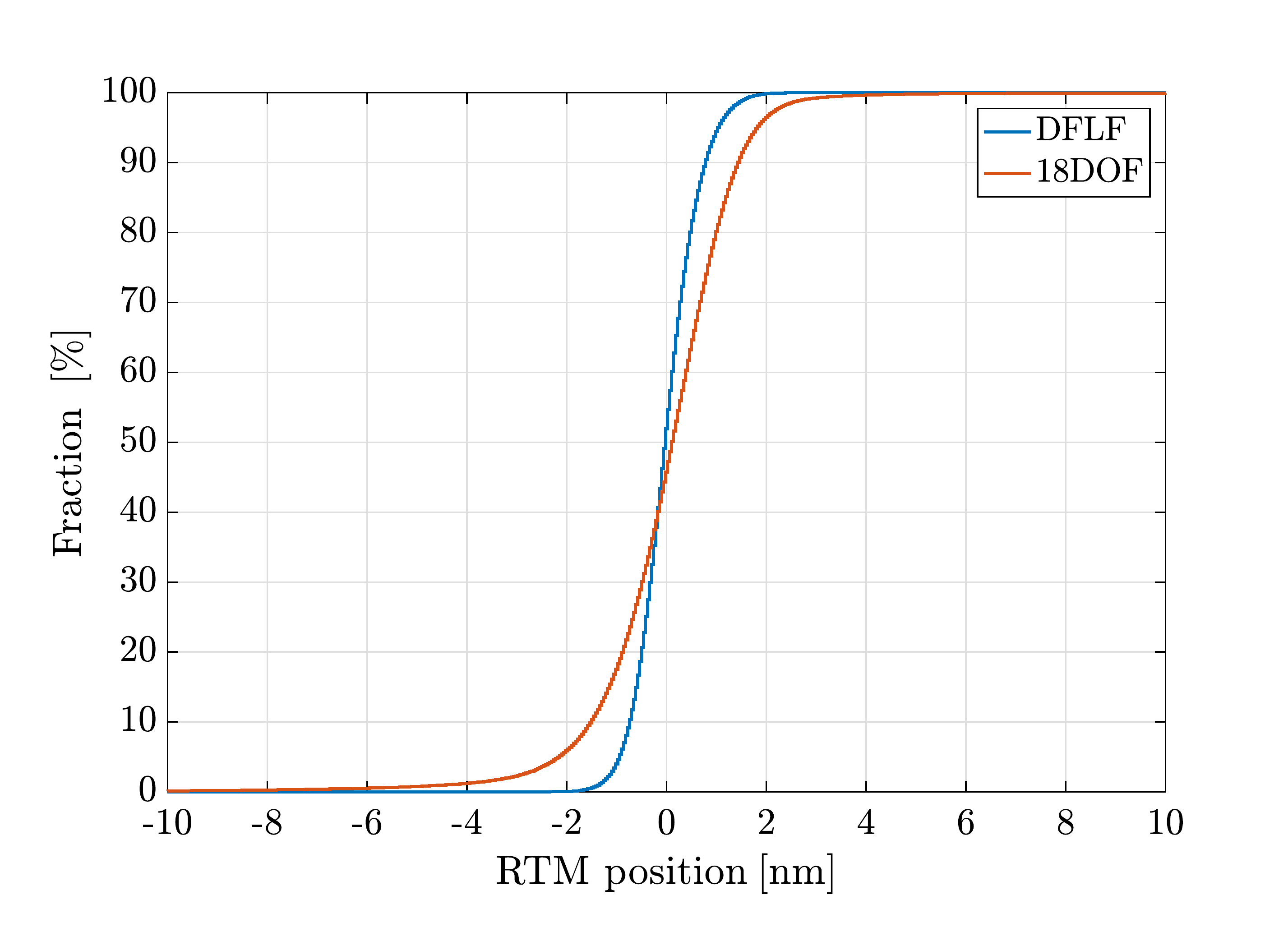}
\caption{Cumulative probability distribution for measured RTM-spacecraft position along the x-direction for the two runs in Figure \ref{fig:position_error_f}. The 95\% confidence intervals for the DFLF and 18DOF position errors are (-1.0,+1.0)$\,$nm and (-2.1,+1.9)$\,$nm respectively.}
 \label{fig:position_error_cdf} 
\label{default}
\end{center}
\end{figure}

Figure \ref{fig:position_error_cdf} plots the measured cumulative probability distribution function for the spacecraft position error along x for the same two runs as are plotted in Figure \ref{fig:position_error_f}. The distributions are well-approximated by a Gaussian distribution and have confidence intervals of (-1.0,+1.0)$\,$nm for DFLF and (-2.1,+1.9)$\,$nm for 18DOF.

\subsection{Differential Acceleration Measurements}
\label{sec:deltag}
\vskip 0.2in
While the primary purpose of DRS operations was to validate the performance of both the drag-free control laws and the CMNT micropropulsion system, a small portion of the operations time, in both the prime and extended missions, was utilized to make differential acceleration measurements of the two test masses. This `$\delta g$' measurement is the primary measurement reported by the LTP collaboration\cite{LPF_PRL_2016, LPF_PRL_2018}.  To a leading approximation, one would not expect a change in the differential acceleration noise when either the control laws or the micropropulsion system was changed. As is extensively discussed in the LTP collaboration publications, the performance of the drag-free system is primarily determined by the physics of the sensor assembly and interferometric readout. For example, the minimum acceleration noise in the $1\,\textrm{mHz}\,\sim10\,\textrm{mHz}$ band is largely determined by the gas pressure around the test mass. The bulk of the LTP operations were composed of experiments to characterize and reduce these various couplings, leading to the improvement in performance from the initial\cite{LPF_PRL_2016} to the final\cite{LPF_PRL_2018} results. 

To first order, a change in the control system does not affect $\delta g$ measurements because the analysis used to construct the $\delta g$ results includes both the error signal (motion of the test mass) as well as the control signal (forces on the test mass).  Second-order effects, such as larger sensitivities to calibration errors or actuation-cross talk, can be present. Indeed, early in the DRS operations, it was noticed that the measured $\delta g$ for Fourier frequencies $\gtrsim 30\,\textrm{mHz}$ was non-stationary and on average higher during DRS operations than during LTP operations. This was traced to the fact that the DRS suspension controller was initially tuned to be `softer' than the corresponding LTP controller, which resulted in larger RMS motion (but less actuation) of the suspended test mass. This larger motion caused an increase in the average sensing noise of the LTP interferometer, which had a known degradation in noise performance if the test masses were allowed to move appreciably from their nominal position.  Once this was understood, the DRS controllers were modified to be `stiffer' at high frequencies, thus reducing the motion of the suspended test mass and recovering the interferometric sensing noise performance observed in the LTP.

One also expects the $\delta g$ results to depend only weakly on which micropropulsion system is used, since the differential nature of the measurement is specifically designed to reject disturbances on the spacecraft platform.  The most direct coupling of microproulsion noise is through test mass `stiffness', which represents the coupling between spacecraft position and force on the test mass, which is physically caused by all of the following: AC electric fields used to control the test masses, stray electrostatic and magnetic fields coupling to test mass charge, and the gradient of the gravitational field due to the spacecraft itself.  This coupling can be reduced by increasing the contribution from the actuation fields so that both test masses have the same stiffness, thus rejecting any coupling of spacecraft motion through the use of `matched' stiffness. In practice, the matched stiffness configuration was not extensively used in either LTP or DRS operations because the intrinsic stiffness of both test mass systems was significantly lower than requirements (and the spacecraft motion due to micropropulsion noise were within requirements).  Micropropulsion noise could also enter the $\delta g$ measurement through more subtle effects such as rotating-frame effects. For example, the measured $\delta g$ signal includes a centrifugal term that arises from the product of the low-frequency rotation of the spacecraft as well as the in-band attitude jitter of the spacecraft. The standard $\delta g$ analysis uses a combination of star-tracker attitude data and test mass torque data to estimate this contribution and subtract it. It is possible that increased angular jitter caused by a noisier microproulsion system could result in a larger contribution that is more difficult to fully subtract.

\begin{figure}
\begin{center}
\includegraphics[width=\columnwidth]{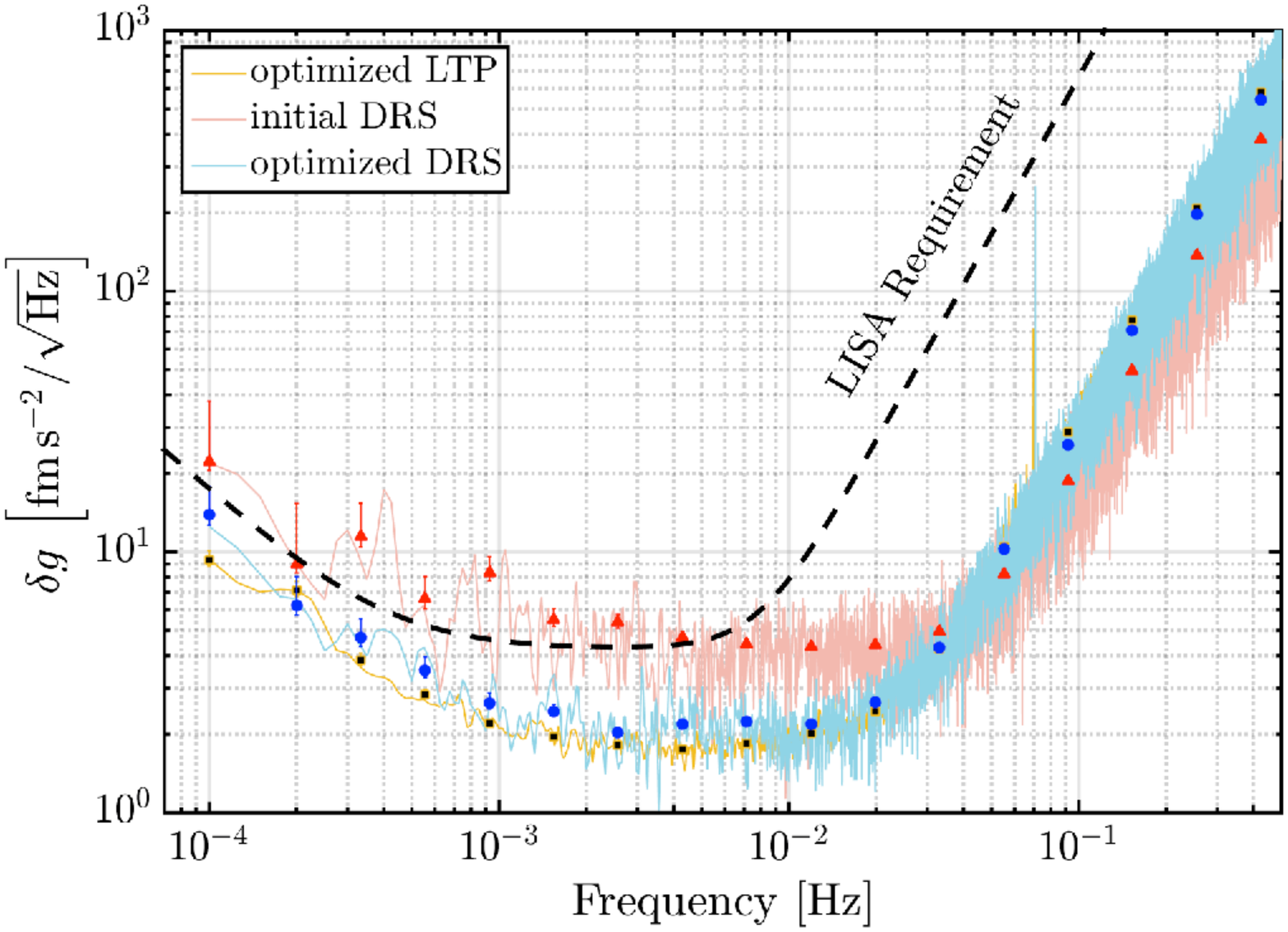}
\caption{Measured residual differential acceleration between Pathfinder's two test masses ($\delta g$) for DRS operations and comparison with LTP configuration. The red trace (initial DRS) is for an all-DRS configuration from early in the mission when all eight CMNTs were operating and the cold gas micropropulsion system was on standby (segment II in Table \ref{tab:segments}). The blue trace (optimized DRS) is for a configuration with the DCS controlling the spacecraft using 7 of 8 CMNTs while four of the cold gas microthrusters provided an open-loop static force (segment IV in Table \ref{tab:segments}). The orange trace is the ultimate published LTP performance\cite{LPF_PRL_2018}.  For all three segments, the solid trace is a linearly-binned amplitude spectral density with a resolution of $250\,\mu$Hz while the solid markers represent logarithmically-binned estimates of the amplitude spectral density with one sigma error bars. The LISA requirements are the single test-mass acceleration requirement as expressed in the 2017 LISA Mission Proposal\cite{LISA_PROPOSAL_2017} with a factor of $\sqrt{2}$ applied to account for the fact that Pathfinder measures differential acceleration between two test masses. \label{fig:delta-g-DRS} }
\label{default}
\end{center}
\end{figure}

Figure \ref{fig:delta-g-DRS} shows our estimate of the amplitude spectral density of $\delta g$ for two DRS configurations as well as the `ultimate' LTP performance\cite{LPF_PRL_2018} and the current estimated requirements for the LISA mission. These data were obtained using the same data analysis pipelines and tools \cite{LTPDA} as used by the LTP collaboration in their major results papers\cite{LPF_PRL_2016, LPF_PRL_2018}.  For all three segments, the solid trace represents the amplitude spectral density obtained using Welch's method of overlapped averaged periodograms of length 40$\,$ks, with a Blackmann-Harris window applied. As was the case for the standard LTP analysis, the lowest reported frequency is the 4th bin (0.1$\,$mHz). The solid points are logarithmically-binned estimates of the amplitude spectral density, including one-sigma error bars. For each time-series segment, the data is reduced in a series of steps that includes estimating the observed acceleration, correcting for the applied force on the non-reference test mass, correcting for stiffness as well as actuation and sensing cross talk, correcting for rotating-frame effects, and removing any impulsive `glitches'.  Each of these steps requires both a model of the underlying contribution to $\delta g$ as well as a set of parameters corresponding to the state of the instrument. Again, to the greatest extent possible, our analysis used an identical set of models and parameters as the corresponding LTP analysis. Similarly, glitches were identified and removed using the same procedure as for the final LTP results. 

The first DRS segment in Figure \ref{fig:delta-g-DRS} (red trace, `initial DRS', data from segment II in Table \ref{tab:segments}) represents the nominal configuration with the DCS controlling the test masses via the LTP and the spacecraft via all eight CMNTs.  Three glitches were identified in this segment, occurring at 2016-10-05 05:50 UTC, 2016-10-04 17:51 UTC, and 2016-10-04 18:07 UTC.  Each glitch was fit and removed using a double-exponential, as described in the final LTP paper\cite{LPF_PRL_2018}. The second DRS segment (blue trace, `optimized DRS', data from segment IV in Table \ref{tab:segments}) represents the optimized DRS performance, obtained after the system had been tuned but also after CMNT \#4 had failed.  Again, after the failure of CMNT \#4, the DCS was modified to control the spacecraft with seven CMNTs in closed-loop as well as an open-loop `crutch' provided by four of LPF's cold gas thrusters.  Two glitches, at 2017-04-23 03:25 UTC and 2017-04-23 13:22 UTC, were identified and removed. The LTP segment (orange trace, `ultimate LTP') is the February 2017 segment plotted in Figure 1 of the final LTP $\delta g$ paper\cite{LPF_PRL_2018}.  

The LISA requirements are the single test-mass acceleration noise requirement from the 2017 LISA Mission Proposal\cite{LISA_PROPOSAL_2017}, but here multiplied by  $\sqrt{2}$  to compensate for the fact that LPF makes a measurement of the differential noise between two test masses whereas the requirement is written for a single test mass.

When comparing the three configurations in Fig. \ref{fig:delta-g-DRS} it is useful to consider three different frequency regimes. At high frequencies ($f\gtrsim 30\,\textrm{mHz}$), the three traces are quite similar, exhibiting a $f^2$ power-law behavior that is caused by white displacement noise in the LTP interferometric readout of the differential test mass position. The most notable differences are the presence of a line feature near 70$\,$mHz in both the LTP and optimized DRS configurations as well as the fact that the initial DRS configuration is slightly lower than the other two. The elevated broad-band displacement noise is likely due to more significant misalignments of the test masses during the extended mission.  During the baseline mission, an extensive campaign was carried out to identify misalignments and actively correct for them by modifying the static offsets of the test mass positions and attitudes. This procedure was shown to significantly reduce the cross-coupling term in the $\delta g$ analysis (see discussion in initial LTP paper\cite{LPF_PRL_2016}). This adjustment was not repeated in the extended mission and it is likely that the offsets may have changed due to either deliberate changes in the spacecraft temperature or other effects such as creep and outgassing. The origin of the 70$\,$mHz line feature is unknown, although the fact that it is not present in the initial DRS configuration, when the cold gas micropropulsion system was placed in standby mode, suggests that it may be related to the cold gas micropropulsion system in some way. 

In the middle band between ($1\,\textrm{mHz}\gtrsim f \gtrsim 10\,\textrm{mHz}$), the three traces show clear differences, with the LTP trace presenting a nearly flat, feature-free noise floor of approximately 1.8$\,\textrm{fm}\,\textrm{s}^{-2} / \sqrt{\textrm{Hz}}$. The initial DRS trace is roughly two times higher at 10$\,$mHz and rises slowly towards lower frequencies. The optimized DRS trace has a lower broad band noise floor than the initial DRS trace, around 3$\,\textrm{fm}\,\textrm{s}^{-2} / \sqrt{\textrm{Hz}}$.  This is still roughly a factor of two higher than the ultimate LTP case, although with the much shorter segment (236 ks vs. 1.15 Ms), the statistics are not as good. As mentioned above, the limiting noise in this band is expected to be gas pressure in the test mass enclosures. After an initial steady decrease in pressure in response to the opening of the vent duct to space, the GRS pressure was primarily controlled by setting the temperature of the GRS housing. The difference in temperatures between the optimized DRS (12.8$\,^{\circ}$C) and the optimized LTP (11.5$\,^{\circ}$C) is not large enough to account for the observed difference in noise,  although the higher noise floor in the initial DRS run, which occurred at a higher temperature (23.5$^{\circ}$C) and several months earlier in the mission, may well be due to increased pressure in the housings.

At the low end of the measured frequency band ($0.1\,\textrm{mHz}\gtrsim f \gtrsim 1\,\textrm{mHz}$), all three traces in Fig. \ref{fig:delta-g-DRS} show a rise with a slope of roughly $f^{-1}$, with their relative amplitudes similar to those in the mid-band. At the lower end of the band, the ultimate DRS noise is slightly higher than that of the ultimate LTP, although the statistics on the DRS measurement are poor. While the long-duration LTP results were able to demonstrate performance down to 20$\,\mu$Hz, a region with important astrophysical implications, the DRS data is not sufficiently long to make any measurements below 0.1$\,$mHz. However, based on the available data it would appear that a drag-free system employing CMNT micropropulsion could meet all of the LISA science requirements while providing significant savings in mass to the flight system.

\section{Summary, Conclusions and Future Work}
\vskip 0.2in
 The ST7-DRS successfully demonstrated NASA developed drag-free control laws and colloid thrusters in space. Colloid thrusters were selected for the mission because of their potential to be used for future mission requiring low thrust noise ($\le 0.1 \mu$ N/ $\sqrt{Hz}$), high precision ($\le 0.1  \mu N$ steps) with a comparatively small amount volume and mass used for propellent. The ST7-DRS controlled the attitude the 424 kg ESA LISA Pathfinder Spacecraft for 103.4 days (~41 days of drag-free control) using less than 1 kg of propellant. The mission was also an example  of applying operational characteristics of colloid thrusters to a drag-free control application.  
 As a technology demonstration, ST7-DRS was a success, meeting its performance requirements. However, the anomalies experienced during the mission would not be acceptable had they occurred in the primary propulsion system of a long-duration mission. Also, the propellent volume must be increased for a several year mission. In the decade since the CMNT thrusters were completed in 2008, significant progress has been made to adapt the technology to a longer mission. The on-orbit performance observed on ST7 is also informing this development and also the planning for the verification and validation testing to demonstrate the improvements have been realized without introducing new problems.
The on-orbit data form ST7 is also being used to infer the performance of colloid thrusters in other applications. While ST7-DRS specifically implemented a drag-free control system, the data collected allows the design of other types of control system. The performance demonstrated on LPF can enable many applications needing ultra-precise pointing, formation flying or dynamic stability, including separated element interferometers, coronagraphs, very large aperture telescopes and fundamental physics experiments. In addition, the control modes and operational approach demonstrated on ST7 are an example as to how to hand over from a higher-noise control system to a very low noise system, and also demonstrated the response to impulse events (and anomalies) that can be used in planning for any of these applications. We look forward to the results of LISA and many other amazing science missions in the future.

\acknowledgments

We would like to acknowledge Dr. Landis Markley for insightful contributions to the design and development of the DRS and DCS.  We acknowledge the significant contributions of Jeff D\'Agostino and Kathie Blackman of the Hammers Co., who implemented and tested the DCS algorithms in the FSW, and who supplied the corresponding simulators to JPL and ESA.  
The data was produced by the NASA Disturbance Reduction System payload, developed under the NASA New Millennium Program and hosted on the LISA Pathfinder mission, which was part of the space-science programme of the European Space Agency. The work of JPL authors was carried out at the Jet Propulsion Laboratory, California Institute of Technology, under contract to the National Aeronautics and Space Administration.  Dr. Slutsky was supported by NASA through the CRESST II cooperative agreement CA 80GSFC17M0002.
The French contribution to LISA Pathfinder has been supported by the CNES (Accord Specific de projet CNES 1316634/CNRS 103747), the CNRS, the Observatoire de Paris and the University Paris-Diderot. E.~Plagnol and H.~Inchausp\'{e} would also like to acknowledge the financial support of the UnivEarthS Labex program at Sorbonne Paris Cit\'{e} (ANR-10-LABX-0023 and ANR-11-IDEX-0005-02).
The Albert-Einstein-Institut acknowledges the support of the German Space Agency, DLR, in the development and operations of LISA Pathfinder. The work is supported by the Federal Ministry for Economic Affairs and Energy based on a resolution of the German Bundestag (FKZ 50OQ0501 and FKZ 50OQ1601). 
The Italian contribution to LISA Pathfinder has been supported  by Agenzia Spaziale Italiana and Istituto Nazionale di Fisica Nucleare.
The Spanish contribution to LISA Pathfinder has been supported by contracts AYA2010-15709 (MICINN), ESP2013-47637-P, and ESP2015-67234-P (MINECO). M.~Nofrarias acknowledges support from Fundacion General CSIC (Programa ComFuturo). F.~Rivas acknowledges an FPI contract (MINECO).
The Swiss contribution to LISA Pathfinder was made possible by the support of the Swiss Space Office (SSO) via the PRODEX Programme of ESA. L.~Ferraioli is supported by the Swiss National Science Foundation.
The UK LISA Pathfinder groups wish to acknowledge support from the United Kingdom Space Agency (UKSA), the University of Glasgow, the University of Birmingham, Imperial College, and the Scottish Universities Physics Alliance (SUPA). @ 2018.  All rights reserved.

\section*{Appendix A: Anomalies experienced during ST7-DRS operations}
\label{sec:ANOMALIES}

The DRS experienced three significant anomalies during the mission that, while they affected operations, did not prevent the mission from meeting its objectives and performance requirements. First, CMNT\#1 demonstrated a reduced maximum current and response time compared to acceptance and thermal vacuum testing prior to launch. One of its nine emitters also blipped on and off between 0.2-0.6 Hz, depending on current level, throughout both the primary and extended missions, significantly increasing thrust noise. Second, the DCIU PROM on cluster 2 suffered a partial memory failure in the location of the thruster control algorithm just after the instrument commissioning was complete, preventing direct thrust commands from being acted on without reseting the DCIU.  Third, near the end of the primary mission, CMNT\#4 experienced a propellant bridge between the emitter and the extractor electrodes, effectively preventing further use of the thruster. This electrical short occurred after the mission performance requirements had been met (including 60 days of operation) and during experiments that were designed to prove performance at more stressful operating conditions. For all three anomalies, solutions or work-arounds were developed to enable operations and continue experiments up to the end of the extended mission.  The next three subsections provide more detail on all three anomalies.

\subsubsection*{CMNT\#1 Performance Reduction} 
\label{sec:anomaly1}
CMNT\#1 in-flight performance was not consistent with its performance in pre-delivery ground acceptance and thermal vacuum (TVAC) ground tests, where it met all response time and thrust noise requirements.  For example, in the TVAC tests, the response time required for CMNT\#1 to increase thrust from 5 to 30 ${\mu}N$ was 9.8 s, which was below the 100 s requirement.  In flight, CMNT\#1 took 2 days longer than the other thrusters to fill the propellant feed system and initially turn on, indicating a significantly reduced maximum propellant flow rate.  After startup and bubble removal in a test during commissioning, it demonstrated a slower response time of about 170 s and a maximum thrust capability that was less than the other thrusters but still within requirements.  This increased response time characteristic continued throughout the mission but would gradually improve and nearly return to meeting the 100 s response time requirement after the microvalve was left open for a number of days.  Whenever CMNT\#1 was deactivated and its microvalve closed, the response time would then decrease again.  Experiments during the extended mission showed that the response time increase was most closely related to how long CMNT\#1's microvalve was closed.  This indicated a problem in the microvalve actuator or a non-constant and not complete blockage in the feed system upstream of the microvalve or in it's flow limiting orifice just upstream of the microvalve seat.  The net result was a requirement to ``prime'' CMNT\#1 by running it in a diagnostic mode at constant current for 30 minutes prior to each start up to improve its response time.  The extra delay in CMNT\#1's response also impacted thrust noise during periods of higher fluctuations in thrust commands (i.e. 18DOF).  Fortunately, the failure of CMNT\#1 to meet this response time requirement did not significantly impact DRS operations during mode transitions or science mode, nor prevent the DRS as a whole from meeting its Level-1 performance requirements.

In flight, CMNT\#1 also experienced beam current spikes of 160-240 nA every 1.75-4 seconds with the frequency and magnitude proportional to the current level throughout the entire mission. The size and high-speed characteristics of the current spikes were consistent with a single emitter turning on and off with $<10$\% duty cycle - an effect colloquially referred to as `blipping'.  During tests designed to allow counting of the number of active emitters on each thruster and from direct measurements of thrust using the GRS, it was clear that CMNT\#1 had one of its nine emitters mostly off. Analyses of these experiments and others suggest that the observed behavior was not due to a bubble but a blockage or constriction in one emitter with a constant hydraulic impedance that impeded the propellant flow rate significantly compared to the other eight emitters.  With a reduced flow rate in just one of the nine emitters, it was not possible to maintain a steady current. 

Both of these performance reductions on CMNT\#1 were observed to start at the same time, during startup, suggesting that they could be related and are both likely caused by an increase in hydraulic resistance; however, the location of the constrictions is not the same.  It should be noted that during startup, a large amount of current $>24 {\mu}A$ was emitted for over 20 minutes, potentially indicating a bubble passing through the microvalve orifice, expanding by a factor of 10 in volume and pushing a significant amount of propellant out of one or more emitter.  This long-duration, high-current event occurred outside of daily real-time observations, fortunately terminating on its own, but potentially damaging one of the emitters in the process.  Follow-on ground tests have shown constrictive damage to emitters that experience abnormally high currents with a build-up of decomposed propellant products, which could explain CMNT\#1's blipping emitter.  A bubble stuck in the microvalve orifice could also explain the reduced flow rate.  The cause of both performance issues with CMNT\#1 continues to be under investigation.  Because the seven other thrusters and microvalves performed nearly the same in flight as on the ground, the CMNT\#1 anomaly could be a yield issue that can be addressed by maturing the technology, improving the microvalve testing, screening, and filling processes and/or flying redundant valves, as is common practice for primary propulsion systems on large science missions.

\subsection*{DCIU PROM Corruption}
\label{sec:anomalyDCIU}
The DCIU on cluster 2 experienced an anomaly on July 8, 2016, immediately after a successful instrument commissioning, resulting in a processor reset and CMTA 2 becoming disabled whenever it received a thrust command. Testing revealed that, while the Thrust Command Mode no longer functioned properly, the Diagnostic Mode and other control routines still functioned properly allowing diagnosis and an eventual work-around.  By design, the DCIU PROM and control software could not be updated on orbit, but fortunately the IAU flight software (FSW) was designed with a back-up thrust control algorithm using the DCIU's Diagnostic Mode in case there was a desire to modify it on orbit.  In this case, the IAU calculated the beam, extractor, and microvalve voltage commands directly from the DCS thrust requests and sent those commands instead of thrust commands to the DCIU.  The IAU FSW and operational sequences required updating and verification to make the embedded thrust control algorithm work properly that enabled this DCIU ``Pass-through Mode'', and operations continued on August 8, 2016. 

The suspected cause of this anomaly is a single radiation event that permanently damaged part of cluster 2's PROM.  While the Pass-Through Mode was a work-around that solved the immediate issue, it also introduced a command and telemetry delay between thrust command processing and beam current and voltage commanding, execution, sensing, and feedback that did not exist when the DCIU had it's own internal control loop functioning properly.  This added delay was 1-2 realtime intervals (RTIs) or 0.1-0.2 seconds, depending on the exact timing between telemetry and command packets passing back and forth between the IAU, DCIUs and on-board computer. As discussed in section \ref{sec:IVnoise}, this extra delay led to a limit-cycle oscillation in the thruster control loop. Fortunately, this limit cycle oscillation was confined to frequencies above the target performance bandwidth and did not impact mission performance. To prevent this anomaly in future missions, a more robust, radiation-hardened, and re-writable EEPROM for the DCIU is recommended.  
 
\subsection*{CMNT\#4 Propellant Bridge}
\label{sec:anomaly4}
Near the end of the primary mission and after all Level 1 mission requirements had been met, CMNT\#4 developed an electrical short (An impedance of $200 M\Omega $, which does constitute a ``short'' in the CMNT system) between the emitter and extractor electrodes due to a suspected propellant bridge, which rendered CMNT\#4 effectively inoperable. The bridge occurred on Oct 27, 2016, after 1670 hours of operation on CMNT\#4. While the exact location of the short on the emitters or extractor electrodes cannot be determined without access to the thruster, neither electrode was shorted to ground potential nor the accelerator electrode, indicating the short was not in the PPU.  The variable nature of the short impedance also indicated that it was an electrically conductive bridge of partially-polymerized propellant formed between the emitter and the extractor, likely after the porous extractor became saturated with propellant from normal and off-nominal operation.  These kind of propellant bridges have been observed on the ground previously due to poorly aligned emitters, saturated porous extractors, or large amounts of excess propellant near the emitter tip prior to operation.

At the time the short occurred, CMNT\#4 was undergoing an experiment to verify the thrust performance model over a wide range of beam voltage and temperature conditions.  At the point of failure, the beam voltage was 4 kV and the thruster temperature was reduced to 20C compared to 6 kV and 25C that are nominal operating conditions.  Operating at a lower beam voltage widened the exhaust beam, increasing the flux of propellant at the edges of the beam onto the extractor electrode.  Operating at reduced temperatures increased the propellant's viscosity and impeded absorption and capillary action of the extractor's pores that are designed to soak-up excess propellant.  It is possible that operating in this off-nominal condition (which was still within the specified operational range) contributed to the failure.

Careful analysis of the ground and in-flight data showed that CMNT\#4 did experience significantly more operation time with bubble-driven flow during ground-based TVAC tests than the other thrusters, which could have increased the flux of propellant to the extractor, filling the pores.  In addition, because of another LPF anomaly unrelated to the DRS that occurred approximately a week before the CMNT\#4 short, all the thrusters were shut down abruptly, which could have allowed some excess propellant to escape out of CMNT\#4's emitters during a thermal transient that caused the propellant to expand without voltage on the electrodes.  The safer and normal version of the thruster shutdown procedure includes multiple steps to decrease the quantity of residual propellant in the emitters, reducing the risk of spraying during periods of prolonged shutdown. Finally, to preserve the stable thermal environment on the spacecraft, all CMNTs were left with PPUs enabled, default voltages on, and microvalves closed during standby mode and station keeping maneuvers, which was not originally specified during operations or tested on the ground.  While this kind of operation should not have caused any additional spray or flux to the extractor, on examining the data more carefully, a $~0.05 {\mu}A$ level current was observed during most of this time in standby mode on CMNT\#4, which was 10 times larger in terms of integrated current or total charge than any other thruster during these same periods, indicating low-level spraying between the electrodes that could have lead to premature saturation of the extractor.

Unfortunately, the ST7 mission did not include a method of measuring the current to the extractor or accelerator electrodes, and since this was not measured except on early engineering model ground testing, it is difficult to quantify how thruster lifetime was impacted for this specific case. Determining how to prevent current flux to the extractor during normal operation, spraying between electrodes with default voltages on during standby mode, and monitoring any current to the extractor and accelerator electrodes, will be critical to the further development of this thruster technology, especially for missions with long lifetime requirements like LISA. Implementing redundant thruster heads will also be important for providing the required lifetimes, as is common practice for the primary propulsion system on large science missions. 

After the CMNT\#4 anomaly and because the locations of the two CMTAs on only the spacecraft x-axis, having just seven operable thrusters gave insufficient rotational authority around the x-axis.  At the end of the nominal mission, with ESA's assistance, a hybrid ``crutch'' mode was developed to continue DRS operations using 4 of the LTP cold gas (CGAS) thrusters to provide a constant thrust bias that replaced what CMNT\#4 would have normally provided. The colloid thruster and cold gas thruster thrust bias levels as well as new operational procedures and sequences were developed and validated on ground-based testbeds for this new operating mode. Both 4 and 2 CGAS thruster configurations were demonstrated on the ST7 testbed; however, the 4 CGAS thruster configuration was preferred to reduce the required colloidal thruster thrust bias levels. This hybrid operation was demonstrated, for all DCS modes, just before the final week of the primary mission,
and it continued successfully through the extended mission. 

\section*{Appendix B: Derivation of response to thrust injections}
\label{sec:mathAppendix}
In this section we present a more detailed derivation of the response of the Pathfinder spacecraft and LTP instrument to the thruster injections that were used to calibrate the CMNTs as described in \ref{sec:thustCal} . 

Let $\hat x_{bf}$, $\hat y_{bf}$, $\hat z_{bf}$ be unit vectors along the body-frame axes of the spacecraft ; let $\vec r_{SC}$ be the position of the spacecraft CoM in some inertial frame; and let $\vec r_{TM1}$ and $\vec r_{TM2}$ be the positions (CoMs) of the two TMs in the same inertial frame. The $x1$  value measured by LTP is $\big(\vec r_{TM1} - \vec r _{SC}\big)\cdot \hat x_{bf}$, and similarly for $x2$.  

Applying Newton's 2nd law to either $x1$ or $x2$, and restricting to the most-sensitive ($\hat x_{bf}$) direction gives
\bea
&\ddot x& \equiv \frac{d^2}{dt^2} \big[\big(\vec r_{TM} -  \vec r_{SC}\big)\cdot \hat x_{bf}\big] \\
&=&   \big[\big(\ddot{\vec r}_{TM} -  \ddot{\vec r}_{SC}\big)\cdot \hat{x}_{bf}\big]  + 2 \big[\big( \,\dot{\vec r}_{TM} -  \dot{\vec r}_{SC}\big)\cdot \dot{\hat x}_{bf}\big] \ \ \ \ \ \    \\
&+& \big[\big(\vec r_{TM} -  \vec r_{SC}\big)\cdot \ddot{\hat x}_{bf}\big]  \, \label{basic} .
\eea 

\noindent Re-arranging the above equation for $\ddot x$ gives
\bea
&F^x_{SC} &= -M_{SC} \ddot x  + \frac{M_{SC}}{M_{TM} }F^x_{TM} \label{b1} \\
&+& 2 M_{SC} \big[\big( \,\dot{\vec r}_{TM} -  \dot{\vec r}_{SC}\big)\cdot \dot{\hat x}_{bf}\big]   \label{b2} \\
& +& M_{SC} \big[\big(\vec r_{TM} -  \vec r_{SC}\big)\cdot \ddot{\hat x}_{bf}\big]  \,  \label{b3} 
\eea
where $F^x \equiv \vec F \cdot \hat x_{bf}$. The first term is Newton's 2nd law in the instrument frame while the later two terms account for the rotation of the frame.To evaluate the rotational frame terms, we define 
\be
\dot{\hat x}_{bf} = \vec \Omega \times \hat x_{bf} \, ,
\ee
where $\vec \Omega$ is the spacecraft 's instantaneous angular velocity, which also implies 
\bea
\ddot{\hat x}_{bf} &=& \dot{\vec \Omega} \times \hat x_{bf} +  \vec \Omega \times \dot{\hat x}_{bf} \, , \\
&=& \dot{\vec \Omega} \times \hat x_{bf} +  \vec \Omega \times  \big(\vec \Omega \times \hat x_{bf} \big)  \\
&=& \dot{\vec \Omega} \times \hat x_{bf} + \big(\vec \Omega \cdot \hat x_{bf}\big) \vec \Omega \, - \, \Omega^2 \, \hat x_{bf} \label{c3} \, .
\eea

The dynamical quantities can be separated into those dominated by the injection, including the TM actuation, and the rest, in order to estimate the sizes and timescales on which they are changing.  If one restricts attention to the thruster force at the injection frequency, then line (\ref{b2}) is
negligible compared to line (\ref{b3}).  Finally, $\ddot{\hat x}_{bf}$ in  (\ref{b3}) is well approximated by the 
$\dot{\vec \Omega} \times \hat x_{bf} $ term in line (\ref{c3}), and terms quadratic in $\Omega$ are negligible.  This implies the approximation
\be
\dot{\Omega}_i = (I^{-1})_{ij}N^j \, .
\ee

If we average the motion of $x1$ and $x2$, the largest rotational effects cancel.  The only rotational effects that do not cancel are proportional to $\Delta z$, defined as the z-displacement of both TMs from the spacecraft CoM, which is on the order of 5$\,$cm. Averaged over the two TMs, and restricting analysis to the injection frequency, all the rotating-frame effects can be approximated by 
\be
- M_{SC} (\Delta z) \dot{\Omega}_y \, 
\ee
or
\be
- M_{SC} (\Delta z)\big[(I^{-1})_{yx}N^x + (I^{-1})_{yy}N^y + (I^{-1})_{yz}N^z \big]\, . 
\ee
Of these, the middle term is by far the largest, but it's simple to carry along the off-diagonal terms.

 If $F_i$ is the amplitude of the thrust from thruster i, then there are matrices $T^{xi}$ and $K^{xi}$ such that the x-component of the force and torque from thruster i are  $T^{xi}F_i$ and $K^{xi}F_i$. These matrices can be derived from the thruster positions and orientations. Plugging these into Eqs.~(\ref{b1})--(\ref{b3}) and re-arranging terms, we arrive at

\be
F_{i|x} =  \frac{1}{2}\big[ -M_{SC} (\ddot x_1 + \ddot x_2) + \frac{M_{SC}}{M_{TM} }(F^x_{TM1} + F^x_{TM2})\big] /(1 + R^x_i)\label{eq:Fresp}
\ee

\noindent where the symbol $F_{i|x}$ denotes "the force exerted by thruster $i$, as estimated from $x$ equation of motion", 
and where the rotational correction term $R^x_i$ is given by
\be
R^x_i = (\Delta z) M_{SC}  \big[ (I^{-1})_{yx} K^{xi}+ (I^{-1})_{yy} K^{yi} + (I^{-1})_{yz} K^{zi} \big] /T^{xi} \, .\label{eq:Rcorrect}
\ee
\\

\section*{Appendix C: Estimates of Thruster and Platform Noise Contributions}
\label{sec:NoiseAppendix}
This appendix provides details of the estimates of contributions to the measured force noise on the spacecraft as listed in Table \ref{tab:SCnoiseComp}. 

\subsection*{CMNT Shot Noise}
CMNT shot noise is an effect of the quantized nature of the electrospray thrust, which is composed of the momentum transfer from discrete droplets. Using a charge to mass ratio of 470 C$/$kg, the current corresponds to a droplet rate of $3 * 10^{15}$ drops/second with an associated shot noise of 0.16 nN when projected into the X direction.  
\subsection*{CMNT Flutter Noise}
Flutter noise refers to variations in the thrust direction, which induce a thrust-dependent thrust noise $S_{flutter} = T\cdot S_{cos\alpha}$ where $\alpha$ is the deviation of the thrust vector from its nominal direction. Ground measurements using a 2-D electrometer array measured $S_\alpha < 10^{-3}\,\textrm{rad}/\sqrt{\textrm{Hz}}$ in the relevant band, which gives $S_{flutter}\sim 0.03\,\textrm{nN}/\sqrt{\textrm{Hz}}$ at the maximum thrust of 30$\,\mu$N.
 
 \subsection*{Solar Force Noise}
 Solar radiation pressure (SRP) produces a force noise along the x-axis that is described by 
 \begin{equation}
 S_{SRP,x} = F_{SRP}\cdot S_{SRP}\cdot\bar{H}+F_{SRP}\cdot S_H,\label{eq:SRPx}
 \end{equation}
 where $F_{SRP}$ is the DC force on the spacecraft due to the SRP, $S_{SRP}$ is the spectral density of the stability of the SRP, $H$ is the angle of the spacecraft about the y-axis and $S_H$ is the spectral density of variations in $H$.  The total DC force on the spacecraft in the anti-Sun direction ($-z$ axis) is measured at $\sim24\,\mu$N. This includes contributions both from the direct solar radiation pressure as well as the differential thermal radiation from the warm sunward side of the spacecraft (radiometer effect). A rough order-of-magnitude estimate is that the the radiometer term is approximately 40\% of the direct term. Hence $F_{SRP}\sim17\,\mu$N. During DRS science operations, $\bar{H}\sim10^{-4}\,$rad, and $S_H\lesssim 10^{-4}\,\textrm{rad}/\sqrt{\textrm{Hz}}$.  Measurements of variations in solar flux give  estimates of $S_{SRP}\lesssim 10^{-3}\,/\sqrt{\textrm{Hz}}$. With these parameters, the second term in (\ref{eq:SRPx}) dominates, with a contribution of $1.7\,\textrm{nN}/\sqrt{\textrm{Hz}}$. 
 
 \subsection*{Radiometer Noise}
 The coupling of the radiometer noise, which is primarily along the z-axis, to motion in the x-axis is the same as for SRP.  Using the same logic as presented for the SRP above, the DC force along z is estimated as $F_{rad}\sim7\,\mu$N and the noise along x due to angular jitter along H is $0.7\,\textrm{nN}/\sqrt{\textrm{Hz}}$. Note that because the coupling mechanism for SRP and Radiometer noise is the same (spacecraft jitter in H), they will add \emph{coherently}. This is taken into account for the noise summations in Table \ref{tab:SCnoiseComp}.
 
\subsection*{Magnetic Field Noise}
 The size of the spacecraft is vastly smaller than the spatial length scale over which the interplanetary $\vec B_{ip}$-field varies, so we treat $\vec B_{ip}$ as spatially uniform over the spacecraft .   The interaction of $\vec B_{ip}$ with the current $\vec j$ in the spacecraft can torque the spacecraft , but it produces no net force. Note that time variations in $\vec B_{ip}$ will cause extra currents to flow in the spacecraft , which create a B-field $\vec B_{sc}$ that partially counteracts the changes in $\vec B_{ip}$ -- this is magnetic shielding--and because magnetic permeability will vary across the spacecraft, the total field  $\vec B_{ip} + \vec B_{sc}$  {\it can} have a significant spatial gradient. But this does not invalidate our earlier argument that there is no net force from the interaction $\vec B_{ip}$ and the total $\vec j$.

The interaction of $\vec B_{ip}$ with the net charge on the spacecraft does produce a net force, which we now estimate. The photo-electric effect  "kicks" electrons off the surface of the solar panels.  Indeed, when the CMNTs (which generate thrust by accelerating positively charged droplets away from the spacecraft) are providing the thrust, these two effects largely cancel, and this is the mechanism that keeps the net charge on the spacecraft small.  The spacecraft potential adjusts until the net spacecraft charging rate averages over time to zero.  The spacecraft potential in equilibrium is $\sim 100$V, or (in cgs units) $\sim \frac{1}{3}$statvolt.  From this, we can estimate the total charge on the spacecraft by $V \sim Q/R$.  Using $R \sim 100$cm, we find $Q \sim 33\ $statcoulomb (or $\approx 10^{-8}\ $ Coulomb).    Using $\vec F_{sc} = c^{-1}  Q \,  \vec v \times \vec B_{ip}$,  $v/c \sim 10^{-4}$, and $B_{ip} \approx 3 \times 10^{-4} \big(\frac{f}{1 mHz}\big)^{-0.8} gauss\, Hz^{-1/2}$, we arrive at $F \sim 10^{-2} nN Hz^{-1/2} \big(\frac{T}{10^7 s}\big)  \big(\frac{f}{1 mHz}\big)^{-0.8} $  (where we have used the conversion $10^{-6}$dyne$ =  10^{-2}$ nN). 

\subsection*{Micrometeoroid Impacts}

The LPF spacecraft occasionally encounters interplanetary dust particles which impart an impulsive momentum to the spacecraft. The size of these particles generally follow a power-law distribution with smaller size particles being more numerous than larger ones. For large impacts, the events can be identified and either subtracted or excised from the data, as was discussed in \ref{sec:CMNTnoiseSC}. For smaller impacts, which are far more numerous, the impulsive momentum may not be recognized as a discrete signal and will instead average out to a force noise.  Using a sample of 44 impact events that were identified in a search of $\sim180$ days of LPF data, a powerlaw estimate of the collision rate $R$, per transferred momentum (in units of $\mu N s$), was estimated to be\footnote{A full analysis of these impacts is the subject of a forthcoming paper}: 
\be
R  =  1.5 \times 10^{-6} \, (\bar p)^{-1.64} \,  s^{-1}\, ,
\ee
where we have defined the dimensionless momentum transfer $\bar p \equiv p/p_0$, with $p_0 = 1 \mu N s $.   Assume that in each collision, the momentum is deposited uniformly over some short time $\delta t$.  (As long as $\delta t$ is short compared to $ 33 s$ [$= 1/(30 mHz)$], we shall see that $\delta t$ drops out of the expression for the force noise spectral density in the measurement band.)  Since the collisions represent shot noise,  force-noise spectrum is some constant $S_0$ up to $f \approx 1/(2\delta t)$ (and falls roughly as $f^{-2}$ at higher $f$), and  $S_0$ times $ (2\delta t)^{-1}$ is the mean-square value of the force from collisions:
\bea
S_0 &=&  (2 \delta t) (p_0)^2 \int_0^{\bar p_t} R(\bar p) \big(\frac{\bar p}{\delta t}\big)^2 (\delta t)  \ d \bar p \\
& = & 2.2 \times 10^{-6} \, \bar p_t^{1.36} (\mu N)^2/Hz  \, ,
\eea
where $\bar p_t$ is some threshold value, above which collisions are individually identified and removed from the data.
Using $\bar p_t \approx 0.5$, and dividing $S^{1/2}_{0}$ by  $\sqrt{3}$ to account for the fact that here we want only the x-component of the force noise, we arrive at
\be
S^{1/2}_{0,x} \approx  0.5 n N\/\sqrt{Hz} \, .
\ee

 \subsection*{Test Mass Force Noise}
Test mass force can by estimated by looking at the measured differential acceleration between the two test masses. As shown in section \ref{sec:deltag}, this is at the $\sim 3\,\textrm{fm}\,\textrm{s}^{-2}/\sqrt{\textrm{Hz}}$ level, which would represent an equivalent spacecraft noise of $\sim 1\,\textrm{pN}/\sqrt{\textrm{Hz}}$.  A limitation of this estimate is that it is not sensitive to common-mode forces on the test masses, such as might be caused by time-varying magnetic fields. However, it seems unlikely that the couplings in each test mass would match sufficiently well to have a 6 order of magnitude difference between the absolute and differential effects. Assuming a worst-case common-mode rejection ratio of $10^3$ gives an upper limit of 1$\,\textrm{nN}/\sqrt{\textrm{Hz}}$.

\end{document}